\documentclass[twocolumn,amsmath,aps,superscriptaddress,showpacs,floatfix,a4paper]{revtex4}

\usepackage{graphicx}
\usepackage{placeins}

\begin{document}

%-----------------------------------------------------------------------
\title{Hydrodynamic interactions and Brownian forces in colloidal suspensions:
 Coarse-graining over time and length-scales}
\author{J.\ T.\ Padding}
\affiliation{Department of Chemistry, Cambridge University,
           Lensfield Road, Cambridge CB2 1EW, UK}
\affiliation{Schlumberger Cambridge Research, High Cross,
           Madingley Road, Cambridge CB3 0EL, UK}
\affiliation{Computational Biophysics,
 University of Twente, PO Box 217, 7500 AE, Enschede, The Netherlands.}
\author{A.\ A.\ Louis}
\affiliation{Department of Chemistry, Cambridge University,
           Lensfield Road, Cambridge CB2 1EW, UK}
\date{\today}
%-----------------------------------------------------------------------

\begin{abstract}

We describe in detail how to implement a coarse-grained hybrid
Molecular Dynamics and Stochastic Rotation Dynamics simulation
technique that captures the combined effects of Brownian and
hydrodynamic forces in colloidal suspensions.  The importance of
carefully tuning the simulation parameters to correctly resolve the
multiple time and length-scales of this problem is emphasized.  We
systematically analyze how our coarse-graining scheme resolves
dimensionless hydrodynamic numbers such as the Reynolds number Re,
which indicates the importance of inertial effects, the Schmidt number
Sc, which indicates whether momentum transport is liquid-like or
gas-like, the Mach number Ma, which measures compressibility effects,
the Knudsen number Kn, which describes the importance of non-continuum
molecular effects and the Peclet number Pe, which describes the
relative effects of convective and diffusive transport. With these
dimensionless numbers in the correct regime the many Brownian and
hydrodynamic time-scales can be telescoped together to maximize
computational efficiency while still correctly resolving the
physically relevant physical processes.  We also show how to control a
number of numerical artifacts, such as finite size effects and solvent
induced attractive depletion interactions.  When all these
considerations are properly taken into account, the measured colloidal
velocity auto-correlation functions and related self diffusion and
friction coefficients compare quantitatively with theoretical
calculations. By contrast, these calculations demonstrate that,
notwithstanding its seductive simplicity, the basic Langevin equation
does a remarkably poor job of capturing the decay rate of the velocity
auto-correlation function in the colloidal regime, strongly
underestimating it at short times and strongly overestimating it at
long times.  Finally, we discuss in detail how to map the parameters
of our method onto physical systems, and from this extract more
general lessons -- keeping in mind that there is no such thing as a
free lunch -- that may be relevant for other coarse-graining schemes
such as Lattice Boltzmann or Dissipative Particle Dynamics.

\end{abstract}


\maketitle

\section{Introduction}

Calculating the {\em non-equilibrium} properties of colloidal
suspensions is a highly non-trivial exercise because these depend both
on the short-time thermal Brownian motion, and the long-time
hydrodynamic behavior of the solvent~\cite{Russ89,Dhon96}. Moreover,
the hydrodynamic interactions are of a many-body character, and cannot
usually be decomposed into a pairwise sum of inter-colloid forces.

The fundamental difficulty of fully including the detailed solvent
dynamics in computer simulations becomes apparent when considering the
enormous {\em time} and {\em length-scale} differences between
mesoscopic colloidal and microscopic solvent particles.  For example,
a typical colloid of diameter $1 \mu \mathrm{m}$ will displace on the order of
$10^{10}$ water molecules!  Furthermore, a molecular dynamics (MD)
scheme for the solvent would need to resolve time-scales on the order
of $10^{-15}$~s to describe the inter-molecular forces, while a colloid
of diameter $1
\mu \mathrm{m}$ in water diffuses over its own diameter in about $1$s.

 Clearly, simulating even an extremely crude molecular model for the
 solvent on the time-scales of interest is completely out of the
 question: some form of {\em coarse-graining} is necessary.  The
 object of this paper is to describe in detail one such scheme.  But
 before we do so, we first briefly discuss a subset of the wide
 variety of different simulation techniques that have been devised to
 describe the dynamics of colloidal suspensions.

  At the simplest level, the effects of
the solvent can be taken into account through Brownian dynamics
(BD)~\cite{Erma75}, which assumes that collisions with the solvent
molecules induce a random displacement of the colloidal particle
positions, as well as a local friction proportional to the their
velocity.  Although, due to its simplicity, Brownian dynamics is
understandably very popular, it completely neglects momentum transport
through the solvent -- as described by the Navier Stokes equations --
which leads to long-ranged hydrodynamic interactions (HI) between the
suspended particles. These HI may fall off as slowly as $1/r$, and can
qualitatively affect the dynamical behavior of the
suspension~\cite{Russ89,Dhon96}.

Beginning with the pioneering work of Ermak and
McCammon~\cite{Erma78}, who added a simple representation of the Oseen
tensor~\cite{Dhon96} to their implementation of BD, many authors have
applied computational approaches that include the HI by an approximate
analytical form.  The most successful of these methods is Stokesian
Dynamics~\cite{Brad88}, which can take into account higher order terms
in a multipole expansion of the HI interactions.  Although some more
sophisticated recent implementations of Stokesian Dynamics have
achieved an ${\cal O}(N \log N)$ scaling with the number of
colloids~\cite{Sier01}, this method is still relatively slow, and becomes
difficult to implement in complex boundary conditions.

Another way to solve for the HI  is by direct numerical
simulation (DNS) methods, where the solid particles are described by
explicit boundary conditions for the Navier-Stokes
equations~\cite{Joseph}.  These methods are better adapted to
non-Brownian particles, but can still be applied to understand the
effects of HI on colloidal dynamics.  The fluid particle dynamics
(FPD) method of Tanaka and Araki~\cite{Tana00}, and a related method
by Yamamoto {\em et al.}~\cite{Naka05}, are two important recent
approaches to solving the Navier Stokes equations that go a step
beyond standard DNS, and simplify the problem of boundary conditions by using a
smoothed step function at the colloid surfaces.  In principle Brownian
motion can be added to these methods~\cite{Tana05}.

\begin{figure}[t]
  \scalebox{0.50}{\includegraphics{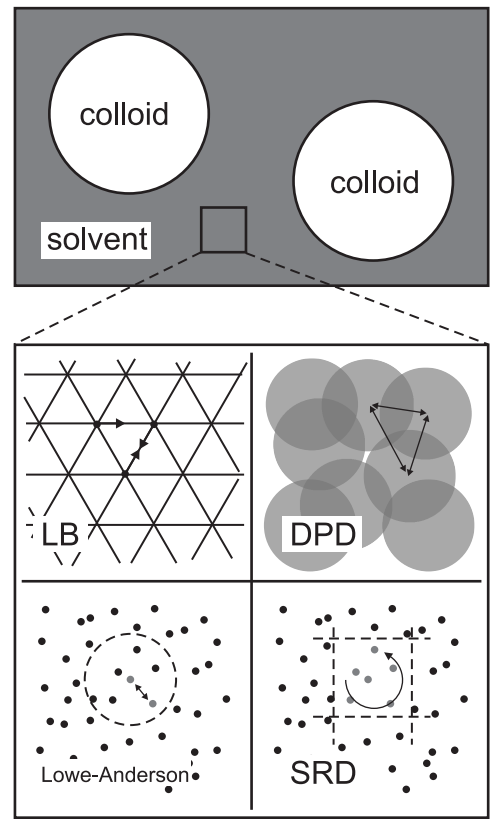}}
\caption{
Schematic picture depicting how a colloidal dispersion can be
coarse-grained by replacing the solvent with a particle-based
hydrodynamics solver such as Lattice Boltzmann (LB), Dissipative
Particle Dynamics (DPD), the Lowe-Anderson thermostat, or Stochastic Rotation
Dynamics (SRD).  Each method introduces an
effective coarse-graining length-scale that is chosen to be smaller than those of the
mesoscopic colloids but much larger than the natural length-scales of
a microscopic solvent.  By obeying local momentum conservation, all
four methods reproduce Navier-Stokes hydrodynamics on larger length
scales. For LB thermal fluctuations must be added in separately, but
these emerge naturally for the other three methods.  In this paper we focus
on SRD, but many of the lessons learned should also apply to other particle 
based coarse-graining methods.
\label{fig:methods} }
\end{figure}

The difficulty of including complex boundary conditions in DNS approaches
has stimulated the use of lattice-gas based techniques to resolve the
Navier Stokes equations.  These methods exploit the fact that only a
few conditions, such as (local) energy and momentum conservation, need
to be satisfied to allow the correct (thermo) hydrodynamics to emerge in the
continuum limit.  Greatly simplified particle collision rules,
easily amenable to efficient computer simulation, are therefore possible, and
complex boundary conditions, such as those that characterize moving
colloids, are easier to treat than in DNS methods.  The most popular
of these techniques is Lattice Boltzmann (LB) where a linearized and
pre-averaged Boltzmann equation is discretized and solved on a
lattice~\cite{Succ01}.  Ladd has pioneered the application of this
method to solid-fluid suspensions~\cite{Ladd93,Ladd01}. This is illustrated 
schematically in Fig.~\ref{fig:methods}. 
 The solid
particles are modeled as hollow spheres, propagated with Newtonian
dynamics, and coupled to the LB solvent through a bounce-back
collision rule on the surface.  The fluctuations that lead to Brownian
motion were modeled by adding a stochastic term to the stress tensor.
Recently this has been criticized and an improved method to include
Brownian noise that does not suffer from some lattice induced
artifacts was suggested~\cite{Cate04}.  A number of other groups have
recently derived alternative ways of coupling a colloidal particle to
a LB fluid~\cite{Loba04,Capu04,Chat05}, in part to simulate the
dynamics of charged systems.

The discrete nature of lattice based methods can also bring
disadvantages, particularly when treating fluids with more than one
length-scale.  Dissipative Particle Dynamics
(DPD)~\cite{Hoog92,Espa95} is a popular off-lattice alternative that
includes hydrodynamics and Brownian fluctuations.  It can be viewed
as an extension of standard Newtonian MD techniques, but with two
important innovations: \\ {\bf 1)} soft potentials that allow large
time-steps and rapid equilibration; \\ {\bf 2)} a Galilean invariant
thermostat that locally conserves momentum and therefore generates the
correct Navier Stokes hydrodynamics in the continuum limit.

In fact, these two methodological advances can be separated.  Soft
potentials such as those postulated for innovation {\bf 1)} may indeed
arise from careful {\em equilibrium} coarse-graining of complex
fluids~\cite{Loui00,Loui01a,Liko01,Klap04}. However, their proper
statistical mechanical interpretation, even for equilibrium
properties, is quite subtle. For example, a potential that generates
the correct pair structure will not normally reproduce the correct
virial pressure due to so called {\em representability
problems}~\cite{Loui02a}.  When used in dynamical simulations, the
correct application of effective potentials is even more difficult to
properly derive. For polymer dynamics, for instance, uncrossability
constraints must be re-introduced to prevent coarse-grained polymers
from passing through one another~\cite{Padd02}. Depletion interactions
in a multi component solution also depend on the relative rates of
depletant diffusion coefficients and particle flow
velocities~\cite{Vlie03,Dzub03}.  At present, therefore, the
statistical mechanical origins of innovation {\bf 1)} of DPD, the use
of soft potentials out of equilibrium, are at best obscure.  We are
convinced that a correct microscopic derivation of the coarse-grained
DPD representation of the dynamics, if this can indeed be done, will
show that the interpretation of such soft potentials depends on
dynamic as well as static (phase-space) averages. Viewing the DPD
particles as static ``clumps'' of underlying fluid is almost certainly
incorrect. It may, in fact, be more fruitful to abandon simple
analogies to the potential energy of a Hamiltonian system, and instead
view the interactions as a kind of coarse-grained
self-energy~\cite{Warr03}.

Innovation {\bf 2)}, on the other hand, can be put on firmer
statistical mechanical footing, see e.g.~\cite{Dunw04}, and can be
usefully employed to study the dynamics of complex systems with other
types of inter-particle interactions~\cite{Sodd03}.  The main
advantage of the DPD thermostat is that, by preserving momentum
conservation, the hydrodynamic interactions that intrinsically arise
from microcanonical MD are preserved for calculations in the canonical
ensemble.  Other thermostats typically screen the hydrodynamic
interactions beyond a certain length-scale~\cite{Warr05}.  For weak
damping this may not be a problem, but for strong damping it could be.

Simulating only the colloids with DPD ignores the dominant solvent
hydrodynamics.  While the solvent could be treated directly by DPD, as
suggested in Fig.~\ref{fig:methods} (see also \cite{Prya05}), this
method is quite computationally expensive because the ``solvent''
particles interact through pairwise potentials.

In an important paper~\cite{Lowe99}, Lowe took these ideas further and
combined the local momentum conservation of the DPD thermostat with the
stochastic nature of the Anderson thermostat to derive a
coarse-grained scheme, now called the Lowe-Anderson thermostat, which
is much more efficient than treating the solvent with full DPD.  It
has recently been applied, for example, to polymer
dynamics~\cite{Lowe05}.

Independently, Malevanets and Kapral~\cite{Male99} derived a method
now called stochastic rotation dynamics (SRD) or also multiple
particle collision dynamics (we choose the former nomenclature here).
In many ways it resembles the Lowe-Anderson thermostat~\cite{Lowe99},
or the much older direct simulation Monte Carlo (DSMC) method of
Bird~\cite{Bird70,Bird94}. For all three of these methods, the
particles are ideal, and move in a continuous space, subject to
Newton's laws of motion. At discrete time-steps, a coarse-grained
collision step allows particles to exchange momentum.  In SRD, space
is partitioned into a rectangular grid, and at discrete time-steps the
particles inside each cell exchange momentum by rotating their
velocity vectors relative to the center of mass velocity of the cell
(see Fig.~\ref{fig:methods}). Similarly, in Lowe's method, a particle
can, with a certain probability, exchange its relative velocity with
another that lies within a certain radius. One can imagine other
collision rules as well. As long as they locally conserve momentum and
energy, just as the lattice gas methods do, they generate the correct
Navier Stokes hydrodynamics, although for SRD it is necessary to
include a grid-shift procedure to enforce Galilean invariance,
something first pointed out by Ihle and Kroll~\cite{Ihle01}.  An
important advantage of SRD is that its simplified dynamics have
allowed the analytic calculation of several transport
coefficients~\cite{Ihle03,Kiku03,Pool04}, greatly facilitating its
use.  In the rest of this paper we will concentrate on the SRD method,
although many of our general conclusions and derivations should be
easily extendible to the Lowe-Anderson thermostat and related methods
summarized in Fig.~\ref{fig:methods}.

SRD can be applied to directly simulate flow, as done for example in
refs~\cite{Lamu01,Alla02}, but its stochastic nature means that a
noise average must be performed to calculate flow lines, and this may
make it less efficient than pre-averaged methods like LB.
Where SRD becomes more attractive is for the simulation of complex
particles embedded in a solvent. This is because in addition to
long-ranged HI, it also naturally contains Brownian fluctuations, and
both typically need to be resolved for a proper statistical mechanical
treatment of the dynamics of mesoscopic suspended particles.  For
example, SRD has been used to study polymers under
flow~\cite{Male00b,Ripo04,Webs05}, the effect of hydrodynamics on 
protein folding and polymer collapse~\cite{Kiku05}, and the
conformations of vesicles under flow~\cite{Nogu05}.  In each of the
previously mentioned examples, the suspended particles are coupled to
the solvent by participating in the collision step.

A less coarse-grained coupling can be achieved  by allowing direct
collisions, obeying Newton's laws of motion, between the SRD solvent
and the suspended particles. Such an approach is important for
systems, such as colloidal suspensions, where the solvent and colloid
time and length-scales need to be clearly separated.  Malevanets and
Kapral~\cite{Male00} derived such a hybrid algorithm that combined a
full MD scheme of the solute-solute and solute-solvent interactions,
while treating the solvent-solvent interactions via SRD. 
 Early
applications were to a two-dimensional many-particle
system~\cite{Inou02}, and to the aggregation of colloidal
particles~\cite{Lee01}.

We have recently extended this approach, and applied it to the
sedimentation of up to 800 hard sphere (HS) like colloids as a
function of volume fraction, for a number of different values of the
Peclet number~\cite{Padd04}.  To achieve these simulations we adapted
the method of Malevanets and Kapral~\cite{Male00} in a number of ways
that were only briefly described in ref.~\cite{Padd04}.  In the current paper we
provide  more in depth analysis of the simulation method we used,
and describe some potential pitfalls.  In particular we focus on the
different time and length-scales that arise from our coarse-graining
approach, as well as the role of  dimensionless hydrodynamic
numbers that express the relative importance of competing physical
phenomena.  Very recently, another similar study of colloidal
sedimentation and aggregation has been carried out~\cite{Hech05}.
Some of our results and analysis are similar, but some are  different,
and we will mention these where appropriate.

After the introductory overview above, we begin in
section~\ref{solvent} by carefully describing the properties of a pure
SRD fluid, focusing on simple derivations that highlight the dominant
physics involved for each transport coefficient. Section~\ref{colloid}
explains how to implement the coupling of a colloidal particle to an
SRD solvent bath, and shows how to avoid spurious depletion
interactions and how to understand lubrication forces.  In
Section~\ref{Dimensionless} we analyze how optimizing the efficiency
of particle based coarse-graining schemes affects different
dimensionless hydrodynamic numbers, such as the Schmidt number, the
Mach number, the Reynolds number, the Knudsen number, and the Peclet
number.  Section~\ref{time-scales} describes the hierarchy of
time-scales that determine the physics of a colloidal suspension, and
compares these to the compressed hierarchy in our coarse-grained
method. In section~\ref{mapping} we tackle the question of how to map
from a coarse-grained simulation, optimized for
computational efficiency, to a real colloidal system.  Finally after a
Conclusions section, we include a few appendices that
discuss amongst other things how to thermostat the SRD system
(Appendix I); why the popular Langevin equation without memory effects
does a remarkably poor job of capturing both the long and the short
time dynamics of the colloidal velocity auto-correlation function
(Appendix II); and, finally, how various physical properties and
dimensionless numbers scale for an SRD simulation, and how these
compare to  a colloid
of radius 10 nm or 1 $\mu$m in H$_2$0 (Appendix III).

\section{Properties of a pure SRD solvent}\label{solvent}

In SRD, the solvent is represented by a large number $N_f$ of
particles of mass $m_f$. Here and in the following, we will call these
``fluid'' particles, with the caveat that, however tempting, they
should not be viewed as some kind of composite particles or clusters
made up of the underlying (molecular) fluid. Instead, SRD should be
interpreted as a Navier Stokes solver that includes thermal noise. The
particles are merely a convenient computational device to facilitate
the coarse-graining of the fluid properties.

\subsection{Simulation method for a pure solvent} 

In the first (propagation) step of the algorithm, the positions and
velocities of the fluid particles are propagated for a time $\Delta
t_c$ (the time between collision steps) by accurately integrating
Newton's equations of motion,
\begin{eqnarray} \label{eq1}
m_f \frac{\mathrm{d}\mathbf{v}_i}{\mathrm{d}t} & = & \mathbf{f}_i\mbox{,} \\
\frac{\mathrm{d}\mathbf{r}_i}{\mathrm{d}t} & = & \mathbf{v}_i\mbox{.} \label{eq2}
\end{eqnarray}
$\mathbf{r}_i$ and $\mathbf{v}_i$ are the position and velocity of
fluid particle $i$, respectively while $\mathbf{f}_i$ is the total
(external) force on particle $i$, which may come from an external
field such as gravity, or fixed boundary conditions such as hard
walls, or moving boundary conditions such as suspended colloids.  The
direct forces between pairs of fluid particles are, however, neglected
in the propagation step.  Herein lies the main advantage -- the
origin of the efficiency -- of SRD.  Instead of directly treating the
interactions between the fluid particles, a coarse-grained collision
step is performed at each time-step $\Delta t_c$: First, space is
partitioned  into cubic cells of volume $a_0^3$. Next,
for each cell, the particle velocities relative to the center of mass
velocity $\mathbf{v}_{\mathrm{cm}}$ of the cell are
rotated:
\begin{equation}\label{eq3}
\mathbf{v}_i \mapsto \mathbf{v}_{\mathrm{cm}} + \mathbf{R}\left(
\mathbf{v}_i - \mathbf{v}_{\mathrm{cm}} \right)\mbox{.}
\end{equation}
$\mathbf{R}$ is a rotation matrix which rotates velocities by a fixed
angle $\alpha$ around a randomly oriented axis. The aim of the
collision step is to transfer momentum between the fluid particles
while conserving the total momentum and energy of each cell.  
Both the collision and the streaming step conserve phase-space volume,
and it has been shown that the single particle velocity distribution
evolves to a Maxwell-Boltzmann distribution~\cite{Male99}.

The rotation procedure can thus be viewed as a coarse-graining of particle
collisions over space {\em and} time. Because mass, momentum, and
energy are conserved locally, the correct hydrodynamic (Navier Stokes)
equations are captured in the continuum limit, \textit{including} the
effect of thermal noise ~\cite{Male99}.

At low temperatures or small collision times $\Delta t_c$, the
transport coefficients of SRD, as originally formulated~\cite{Male99},
show anomalies caused by the fact that fluid particles in a given cell
can remain in that cell and participate in several collision
steps~\cite{Ihle01}. Under these circumstances the assumption of
molecular chaos and Galilean invariance are incorrect.  However, this
anomaly can be cured by applying a random shift of the cell
coordinates before the collision step~\cite{Ihle01,Ihle03}.

It should also be noted that the collision step in SRD does not
locally conserve angular momentum.  As a consequence, the stress
tensor $\mathbf{\sigma}$ is not, in general, a symmetric function of the
derivatives of the flow field (although it is still rotationally
symmetric)~\cite{Pool04}.  The asymmetric part can be interpreted as a
viscous stress associated with the vorticity $\nabla \times
\mathbf{v}$ of the velocity field $\mathbf{v}(\mathbf{r})$.  The stress tensor will
therefore depend on the amount of vorticity in the flow field.
However, the total force on a fluid element is determined not by the
stress tensor itself, but by its divergence $\nabla \cdot \mathbf{\sigma}$,
which is what enters the Navier Stokes equations.  Taking the
divergence causes the explicit vorticity dependence to drop out (the
gradient of a curl is zero).  In principle, the stress tensor could be
made symmetric by applying random rotations as well as random
translations to the cell.  Or, alternatively, angular momentum can be
explicitly conserved by dynamically adapting the collision rule, as
done by Ryder {\em et al}~\cite{Ryde05}, who found no significant
differences in fluid properties (although there is, of course, less
flexibility in choosing simulation parameters).  This result has
been confirmed by some recent theoretical calculations~\cite{Ihle05},
that demonstrate that the asymmetry of the stress tensor has only a
few consequences, such as a correction to the sound wave attenuation
associated with viscous dissipation of longitudinal density waves.
These are not important for the fluid properties we are trying to
model, and so, on balance, we chose not to implement possible fixes
to improve on angular momentum conservation.

\subsection{Transport coefficients and the dimensionless mean-free path}

The simplicity of SRD collisions has facilitated the analytical
calculation of many transport
coefficients~\cite{Kiku03,Pool04,Ihle03,Ihle05}.  These analytical
expressions are particularly useful because they enable us to
efficiently tune the viscosity and other properties of the fluid,
without the need for trial and error simulations.  In this section we
will summarize a number of these transport coefficients, where possible
giving a simple derivation of the dominant physics.

\begin{table}
\begin{center}
\caption{Units and simulation parameters for an SRD fluid.  The parameters
listed in the table all need to be  fixed independently to determine a
simulation.
\label{table:units}
}
\begin{ruledtabular}
\begin{tabular}{l}

\begin{tabular}{l|l}
\begin{tabular}{l}
\begin{tabular}{l}
Basic Units
   \\ \hline \noalign{\medskip}
  \end{tabular}  \\ \medskip

  \begin{tabular}{ll} 
$a_0$ &= length \\  \noalign{\medskip}
$k_B T$ & = energy \\  \noalign{\medskip}
$m_f$ &= mass\\  \noalign{\medskip}
 & \\  \noalign{\medskip}
 & \\  \noalign{\medskip}
 & \\  \noalign{\medskip}

\end{tabular} \\ \noalign{\medskip}

\end{tabular} &
\begin{tabular}{l}

\begin{tabular}{l}
Derived Units
   \\ \hline \noalign{\medskip}
  \end{tabular}  \\ \medskip

  \begin{tabular}{l} 
$\displaystyle t_0 = a_0 \sqrt{\frac{m_f}{k_B T}}$ \hspace*{1cm} = time \\  \noalign{\medskip}

$\displaystyle D_0 = \frac{a_0^2}{t_0} = a_0 \sqrt{\frac{k_B T}{m_f}}$
 = diffusion constant \\  \noalign{\medskip}

$\displaystyle \nu_0 = \frac{a_0^2}{t_0} = a_0 \sqrt{\frac{k_B T}{m_f}}$  = kinematic viscosity \\  \noalign{\medskip}

$\displaystyle \eta_0 = \frac{\gamma m_f}{t_0 a_0} = \frac{\sqrt{m_f k_B T}}{a_0^2}$ = viscosity\\  \noalign{\smallskip}

\end{tabular} 

\end{tabular}

\end{tabular}

\\
\hline 
\noalign{\medskip}
\begin{tabular}{l}
Independent fluid simulation parameters 
   \\ \hline \noalign{\medskip}
  \end{tabular}  \\ \medskip

 \begin{tabular}{ll} 
$\gamma$ &= average number of particles per cell \\  \noalign{\medskip}
$\Delta t_c$ & = SRD collision time step \\  \noalign{\medskip}
$\alpha$ &= SRD rotation angle\\  \noalign{\medskip}
$L$ &= box length\\  \noalign{\medskip}
\end{tabular} \\
\hline \noalign{\medskip}

\begin{tabular}{l}
Independent colloid simulation parameters
   \\ \hline \noalign{\medskip}
  \end{tabular}  \\ \medskip

 \begin{tabular}{ll} 
$\Delta t_{MD}$ & = MD  time step \\  \noalign{\medskip}
$\sigma_{cc}$ &= colloid-colloid collision diameter \,\, Eq.~(\ref{eq:col-col}) 
\\  \noalign{\medskip}
$\epsilon_{cc}$ &= colloid-colloid energy scale \,\, Eq.~(\ref{eq:col-col}) 
\\  \noalign{\medskip}
$\sigma_{cf}$ &= colloid-fluid collision diameter \,\, Eq.~(\ref{eq:col-fluid}) 
\\  \noalign{\medskip}

$\epsilon_{cf}$ &= colloid-fluid energy scale \,\, Eq.~(\ref{eq:col-fluid}) 
\\  \noalign{\medskip}

$N_c$ & = number of colloids \\ \noalign{\medskip}
$M_c$ & =  colloid mass \\ \noalign{\medskip}
\end{tabular} \\

\end{tabular}
\end{ruledtabular}
\end{center}
\end{table}

\subsubsection{Units and the dimensionless mean-free path}

 In this paper we will use the following units: lengths will be in
units of cell-size $a_0$, energies in units of $k_B T$ and masses in
units of $m_f$ (This corresponds to setting $a_0=1$, $k_BT=1$ and
$m_f=1$). Time, for example, is expressed in units of $t_0 =a_0
\sqrt{m_f/k_BT}$, the number density $n_f = \gamma/a_0^3$  and other derived units can be found in table~\ref{table:units}.
 We find it instructive to express the transport
coefficients and other parameters of the SRD fluid in terms of the
dimensionless mean-free path
\begin{equation}\label{eq:mean-free-path}
\lambda = \frac{\Delta t_c}{a_0} \sqrt{\frac{k_B
T}{m_f}} =  \frac{\Delta t_c}{t_0},
\end{equation}
 which provides a measure of the average
fraction of a cell size that a fluid particle travels between
collisions.

This particular choice of units helps 
highlight  the basic physics of the coarse-graining
method.  The (nontrivial) question of how to map them on to the units
of real physical system will be discussed in section~\ref{mapping}.

\subsubsection{Fluid self-diffusion constant}

A simple back of the envelope estimate of the self-diffusion constant
$D_f$ of a fluid particle can be obtained from a random-walk picture.
In a unit of time $t_0$, a particle will experience $1/\lambda$
collisions, in between which it moves an average distance $\lambda
a_0$.  Similarly, a heavier (tagged) particle of mass $M_t$, which
exchanges momentum with the fluid by participating in the
coarse-grained collision step, will move an average distance $l_M
= \Delta t_c \sqrt{k_B T/M_t} 
 =
\lambda a_0/\sqrt{M_t}$ between collisions. By viewing this motion as 
a random walk of step-size $l_M$, the diffusion coefficient follows:
\begin{equation}\label{eq:diffusion-1}
 \frac{D}{D_0} \sim \frac{m_f}{M_t} \lambda,
\end{equation}
expressed in units of $D_0 = a_0^2/t_0 = a_0
\sqrt{k_B T/m_f}$. 
  The diffusion coefficient
$D_f$ for a pure fluid particle of mass $m_f$ is therefore given by
$D_f/D_0 \approx \lambda$.

A more systematic derivation of the diffusion coefficient of a fluid
particle, but still within a random collision approximation, results in the
following expression~\cite{Ihle03,Ripo05}:
\begin{equation}\label{eq:diffusion}
\frac{D_f}{D_0} = \lambda \left[\frac{3}{2 (1 - \cos(\alpha))}\left(\frac{\gamma}{\gamma -1}\right) - \frac{1}{2}\right].
\end{equation}
The dependence on $\gamma$ is weak. If, for example, we take
$\alpha=\pi/2$, the value used in this paper, then $\lim_{\gamma
\rightarrow \infty} D_f/D_0 = \lambda$, the same as Eq.(\ref{eq:diffusion-1}).

A similar expression can be derived for the self-diffusion coefficient
of a heavier tagged particle of mass $M_t$~\cite{Ihle03,Ripo05}:
\begin{equation}\label{eq:tagged}
\frac{D_t}{D_0} = \frac{\lambda m_f}{M_t} \left[\frac{3}{2 (1 - \cos(\alpha))}\left(\frac{\gamma + \frac{M_t}{m_f}}{\gamma}\right) - \frac{1}{2}\right].
\end{equation}
Note that this equation does not quite reduce to
Eq.~(\ref{eq:diffusion}) for $M_t/m_f =1$ (because of slightly different approximations), but the relative discrepancy decreases
with increasing $\gamma$. 

While Eq.~(\ref{eq:diffusion}) is accurate for larger mean-free paths,
where the random collision approximation is expected to be valid, it
begins to show deviations from simulations for $\lambda \lesssim
0.6$~\cite{Ripo05}, when longer-time kinetic correlations begin to
develop.  Ripoll {\em et al}~\cite{Ripo05} argue that these
correlations induce interactions of a hydrodynamic nature that enhance
the diffusion coefficient for a fluid particle.  For example, for
$\lambda = 0.1$ and $\alpha=\frac34 \pi$, they measured a fluid
self-diffusion constant $D_f$ that is about $25\%$ larger than the
value found from Eq.~(\ref{eq:diffusion}). The enhancement is even
more pronounced for heavier particles: for the same simulation
parameters they found that $D_t$ was enhanced by about $75\%$ over the
prediction of Eq.~(\ref{eq:tagged}) when $M_t \gtrsim 10$.

We note that coupling a large particle to the solvent through
participation in the coarse-grained collision step leads to a
diffusion coefficient which scales with mass as $D_t/D_0 \sim 1/M_t$,
whereas if one couples a colloid of radius $R_c$ to the solvent
through direct MD collisions, one expects $D_t/D_0 \sim a_0/R_c \sim
(m_f/M_t)^{\frac13}$.
 Moreover, it has been shown ~\cite{Ripo05} that the effective
hydrodynamic particle radius is approximately given by $a \approx 2a_0/(\pi \gamma)$.
For any reasonable average number of fluid particles per cell, the effective
hydrodynamic radius $a$ of the heavier particle is therefore much less than
the collision cell size $a_0$. On the other hand, the hydrodynamic field is
accurately resolved down to a scale comparable to $a_0$
({\em vide infra}). What this implies is that this coupling method will yield
 correct hydrodynamic interactions {\it only at large distances}, when the colloids are
more than several hydrodynamic radii apart.
All the above suggests that some care must be taken when
interpreting the dynamics of heavier particles that couple through the
coarse-grained collision step,  especially when two or more heavy
particles are in close proximity.

\subsubsection{Kinematic viscosity} 
The spread of a velocity fluctuation $\delta {\bf v}({\bf r})$ in a 
fluid can be described by a diffusion equation~\cite{Hydrobook}:
\begin{equation}\label{eq:kinematic_diffusion}
\frac{ \partial \delta {\bf v}({\bf r})}{\partial t} = \nu \nabla^2 \delta {\bf v}({\bf r}) 
\end{equation}
where $\nu$ is the kinematic viscosity, which determines the rate at
which momentum or vorticity ``diffuses away''.  The units of kinematic
viscosity are $\nu_0 = a_0^2/t_0= a_0\sqrt{ k_B T/m_f}$ which are the
same as those for particle self diffusion i.e.\ $D_0 = \nu_0$.

Momentum is transported through two mechanisms:

\noindent {\bf 1)} By particles streaming between collision steps, leading to a ``kinetic'' contribution to the kinematic viscosity
$\nu_{kin}$.  Since for this gas-like contribution the momentum is
transported by particle motion, we expect $\nu_{kin}$ to scale like
the particle self-diffusion coefficient $D_f$, i.e.\ $\nu_{kin}/\nu_0
\sim \lambda$.

\noindent {\bf 2)} By momentum being re-distributed among the particles of
each cell during the collision step, resulting in  a ``collisional''
contribution to the kinematic viscosity $\nu_{\mathrm{col}}$.  This mimics the
way momentum is transferred due to inter-particle collisions, and
would be the dominant contribution in a dense fluid such as water at
standard temperature and pressure. Again a simple random-walk argument
explains the expected scaling in SRD: Each collision step distributes momentum among
particles in a cell, making a step-size that scales like $a_0$.  Since
there are $1/\lambda$ collision steps per unit time $t_0$, this
suggests that the collisional contribution to the kinematic viscosity should scale as $\nu_{col}/\nu_0 \sim 1/\lambda$.

Accurate analytical expressions for the kinematic viscosity $\nu =\nu_{kin} +  \nu_{\mathrm{col}}$  of SRD
have been derived~\cite{Kiku03,Ihle05}, and these can be 
rewritten in the following dimensionless form:
\begin{eqnarray}\label{eq:nukin}
\frac{\nu_{\mathrm{kin}}}{\nu_0}
 & = & \frac{\lambda}{3} \,\, f_{\mathrm{kin}}^\nu(\gamma,\alpha)\\
\frac{\nu_{\mathrm{\mathrm{col}}}}{\nu_0} & = & \frac{1}{18 \lambda}
 \,\,f_{\mathrm{\mathrm{col}}}^\nu(\gamma,\alpha)\mbox{.}\label{eq:nucol}
\end{eqnarray} where 
the dependence on the collisional angle $\alpha$ and fluid number
density $\gamma$ is subsumed in the following two factors:
\begin{equation}\label{eq:fkin}
f_{\mathrm{kin}}^\nu(\gamma,\alpha)= \frac{15 \gamma}{(\gamma - 1 +
\mathrm{e}^{-\gamma}) (4-2 \cos (\alpha) - 2 \cos (2 \alpha))}
-\frac{3}{2}
\mbox{,}
\end{equation}
\begin{equation}\label{eq:fcol}
f_{\mathrm{\mathrm{col}}}^\nu(\gamma,\alpha)  = 
 (1 - \cos (\alpha))
(1 - 1/\gamma + \mathrm{e}^{-\gamma}/\gamma)\mbox{.}
\end{equation}
These factors only depend weakly on $\gamma$ for the typical
parameters used in simulations.  For example, at $\alpha=\pi/2$ and
$\gamma=5$, the angle and number density we use in this paper, they
take the values $f_{\mathrm{kin}}^\nu (5,\pi/2)= 1.620$ and
$f_\mathrm{\mathrm{col}}^\nu (5,\pi/2) = 0.801$. For this choice of collision
angle $\alpha$, they monotonically converge to $f_{\mathrm{kin}}=
f_{\mathrm{\mathrm{col}}}=1$ in the limit of large $\gamma$.

\subsubsection{Shear viscosity}

Suppose the fluid is sheared in the x direction, with the gradient of the
average flow field in the y direction. Two neighboring fluid elements
with different y-coordinates will then experience a friction force in the
x-direction, as expressed by the xy component of the stress
tensor $\mathbf{\sigma}$, which for a simple liquid is linearly proportional
to the instantaneous flow field gradient,
\begin{equation}\label{eq:stress-tensor}
\sigma_{xy} = \eta \frac{\partial v_x(y)}{\partial y}\mbox{.}
\end{equation}
The coefficient of proportionality $\eta$ is called the
shear viscosity, and is it related to the kinematic viscosity by $\eta
= \rho_f \nu$, where $\rho_f = m_f \gamma/a_0^3$ is the fluid mass
density.  From Eqs.~{\ref{eq:nukin}--\ref{eq:fcol}} it follows that
the two contributions to the shear viscosity can be written in
dimensionless form as:
\begin{eqnarray}\label{eq:etakin}
\frac{\eta_{\mathrm{kin}}}{\eta_0}
 & = & \frac{\lambda \gamma}{3} \,\, f_{\mathrm{kin}}^\nu(\gamma,\alpha)\\
\frac{\eta_{\mathrm{\mathrm{col}}}}{\eta_0} & = & \frac{\gamma}{18 \lambda}
 \,\,f_{\mathrm{\mathrm{col}}}^\nu(\gamma,\alpha)\mbox{.}\label{eq:etacol}
\end{eqnarray}
where $\eta_0 = m/a_0t_0= \sqrt{ m_f k_B T}/a_0^2$ is the unit of 
shear viscosity.

In contrast to the expressions for the diffusion of a fluid particle
or a tagged particle, Eqs.~(\ref{eq:nukin})~-~(\ref{eq:etacol})
compare quantitatively to simulations over a wide range of
parameters~\cite{Kiku03,Ihle03,Ripo05}. 
 For the
parameters we used in our simulations in ~\cite{Padd04,Padd05}, i.e.\
$\lambda = 0.1, \alpha=\pi/2, \gamma=5$, the collisional contribution
to the viscosity dominates: $\nu_{kin} = 0.054 \nu_0$ and $\nu_{\mathrm{col}} =
0.45 \nu_0$.  This is typical for $\lambda \ll 1$, where $\nu$ and
$\eta$ can be taken to a good first approximation by the collisional
contribution only.  In fact throughout this paper we will mainly focus
on this small $\lambda$ limit.

It is also instructive to compare the expressions derived in this
section to what one would expect for simple gases, where, as famously
first derived and demonstrated experimentally by Maxwell in the 1860's
\cite{Maxwell1860},
the shear viscosity is independent of density. This
result differs from the kinetic (gas-like) contribution to the
viscosity in Eq.~(\ref{eq:etakin}), because in a real dilute gas the
mean-free path scales as $\lambda \propto 1/\gamma$, canceling the
dominant  density dependence in $\eta_{kin} \propto \gamma
\lambda$.  The same argument explains why the self-diffusion and
kinematic viscosity of the SRD fluid are, to first order, independent
of $\gamma$, while in a gas they would scale as $1/\gamma$.  In SRD,
the mean-free path $\lambda$ and the density $\gamma$ can be varied
independently. Moreover, the collisional contribution to the viscosity
adds a new dimension, allowing a much wider range of physical fluids
to be modeled than just simple gases.

\section{Colloid simulation method}\label{colloid}

Malevanets and Kapral~\cite{Male00} first showed how to implement a
hybrid MD scheme that couples a set of colloids to a bath of SRD
particles.  In this section, we expand on their method, describing in
detail the implementation we used in ref.~\cite{Padd04}.  We restrict
ourselves to HS like colloids with steep interparticle repulsions,
although attractions between colloids can easily be added on.  The
colloid-colloid and colloid-fluid interactions, $\varphi_{cc}(r)$ and
$\varphi_{cf}(r)$ respectively, are integrated via a normal MD
procedure, while the fluid-fluid interactions are coarse-grained with
SRD.  Because the number of fluid particles vastly outnumbers the
number of HS colloids, treating their interactions approximately via
SRD greatly speeds up the simulation.

\subsection{Colloid-colloid and colloid-solvent interactions}

Although it is possible to implement an event-driven dynamics of HS
colloids in an SRD solvent~\cite{Wyso05}, here we approximate pure HS
colloids by  steep repulsive interactions of the WCA form~\cite{Hans86}:
\begin{equation}\label{eq:col-col}
\varphi_{cc}(r) = \left\{
\begin{array}{ll}
	4 \epsilon_{cc} \left[ \left(\displaystyle \frac{\sigma_{cc}}{r}\right)^{48} - \left(\displaystyle \frac{\sigma_{cc}}{r}\right)^{24} +\displaystyle \frac{1}{4} \right] & (r\leq 2^{1/24}\sigma_{cc}) \\
	0 & (r> 2^{1/24}\sigma_{cc}).
\end{array}
\right.
\end{equation}
Similarly, the colloid-fluid interaction takes the WCA form:
\begin{equation}\label{eq:col-fluid}
\varphi_{cf}(r) = \left\{
\begin{array}{ll}
	4 \epsilon_{cf} \left[ \left(\displaystyle \frac{\sigma_{cf}}{r}\right)^{12} - \left(\displaystyle \frac{\sigma_{cf}}{r}\right)^6 + \displaystyle \frac{1}{4} \right] & (r\leq 2^{1/6}\sigma_{cf}) \\
	0 & (r> 2^{1/6}\sigma_{cf}).
\end{array}
\right.
\end{equation}
 The   mass $m_f$ of a fluid
particle is typically much smaller than the mass $M_c$ of a colloid, so that
the average thermal velocity of the fluid particles is larger than that of the colloid particles by a factor
$\sqrt{M_c/m_f}$. For this reason the time-step $\Delta
t_{\mathrm{MD}}$ is usually restricted by the fluid-colloid
interaction~(\ref{eq:col-fluid}),  allowing fairly large exponents $n$
for the colloid-colloid interaction $\varphi_{cc}(r) = 4
\epsilon_{cc} \left[ (\sigma_{cc}/r)^{2n} - (\sigma_{cc}/r)^n + 1/4
\right]$.  We choose $n=24$, which makes the colloid-colloid potential  steep, more like hard-spheres, while still soft enough to allow the time-step to be set by the colloid-solvent interaction.

The positions and velocities of the colloidal spheres  are
propagated through the Velocity Verlet algorithm
~\cite{AllenTildesley} with a time step $\Delta t_{\mathrm{MD}}$:
\begin{eqnarray}\label{Verlet}
\mathbf{R}_i\left( t + \Delta t_{\mathrm{MD}} \right) &=& \displaystyle   \mathbf{R}_i\left( t \right)
+ \mathbf{V}_i\left( t \right) \Delta t_{\mathrm{MD}} + \frac{\mathbf{F}_i\left( t \right)}{2M_c}
\Delta t_{\mathrm{MD}}^2, \,\, \\
\mathbf{V}_i\left( t + \Delta t_{\mathrm{MD}} \right) & = & \displaystyle \mathbf{V}_i\left( t \right) +
\frac{\mathbf{F}_i\left( t \right) + \mathbf{F}_i\left( t + \Delta t_{\mathrm{MD}}\right)}{2M_c}
\Delta t_{\mathrm{MD}}. \,\, \nonumber \\
\end{eqnarray}
$\mathbf{R}_i$ and $\mathbf{V}_i$ are the position and velocity of
colloid $i$, respectively.  $\mathbf{F}_i$ is the total force on that
colloid, exerted by the fluid particles, an external field, such as
gravity, external potentials such as repulsive walls, as well as other
colloids within the range of the interaction
potential~(\ref{eq:col-col}).

  The positions ${\bf r}_i$ and velocities ${\bf v}_i$ of SRD
  particles are updated by similarly solving Newton's
  Eqns.~(\ref{eq1},\ref{eq2}) every time-step $\Delta t_{MD}$, and
  with the SRD procedure of Eq.~(\ref{eq3}) every time-step $\Delta
  t_c$.

Choosing both $\Delta t_{MD}$ and $\Delta t_c$ as large as possible
enhances the efficiency of a simulation.  To first order, each
timestep is determined by different physics $\Delta t_{MD}$ by the
steepness of the potentials, and $\Delta t_c$ by the desired fluid
properties, and so there is some freedom in choosing their relative
values. We used $\Delta t_c/\Delta t_{\mathrm{MD}} =4$ for our
simulations of sedimentation~\cite{Padd04}, but other authors have
used ratios of $50$~\cite{Male00} or even an order of magnitude larger
than that~\cite{Hech05}.  Later in the paper we will revisit this
question, linking the time-steps to various dimensionless hydrodynamic
numbers and Brownian time-scales.

\subsection{Stick and slip boundary conditions}

\begin{figure}[t]
  \scalebox{0.50}{\includegraphics{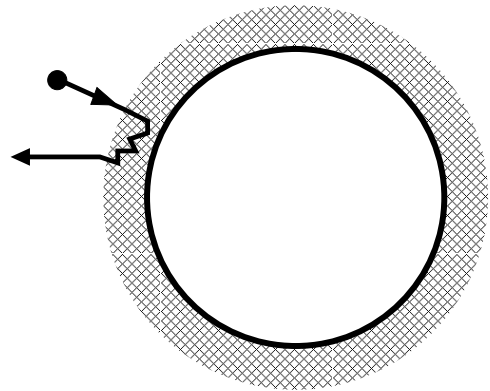}}
\caption{Schematic picture depicting how a fluid molecule interacts with 
a colloid, imparting both linear and angular momentum.  Near the
colloidal surface, here represented by the shaded region, there may be
a steric stabilization layer, or a double-layer made up of co- and
counter-ions.  In SRD, the detailed manner in which a fluid particle
interacts with this boundary layer is represented by a coarse-grained
stick or slip boundary condition.
\label{fig_stickboundary}
}
\end{figure}

Because the surface of a colloid is never perfectly smooth, collisions
with fluid particles transfer angular as well as linear momentum. 
As demonstrated in Fig.~\ref{fig_stickboundary}, the
exact molecular details of the colloid-fluid interactions may be very
complex, and mediated via co- and counter-ions, grafted polymer
brushes etc\ldots.  However, on the time and length-scales over which
our hybrid MD-SRD method coarse-grains the solvent, these interactions
can be approximated by {\em stick boundary conditions}: the tangential
velocity of the fluid, relative to the surface of the colloid, is zero
at the surface of the colloid~\cite{Bren83,Bocq94}.  For most
situations, this boundary condition should be sufficient, although in
some cases, such as a non-wetting surface, large slip-lengths may
occur~\cite{Barr99}.

 In computer simulations, stick boundary conditions may be implemented
 by bounce-back rules, where both parallel and perpendicular
 components of the relative velocity are reversed upon a collision with a
 surface.  These have been applied by Lamura {\em et
 al}~\cite{Lamu01}, who needed to modify the bounce-back rules
 slightly to properly reproduce stick boundaries for a Poiseuille flow
 geometry.

Stick boundaries can also be modeled by a stochastic rule.  After a
collision, the relative tangential velocity $v_t$ and relative normal velocity $v_n$ are
taken from the distributions:
\begin{eqnarray}
\label{eq_stochastic1}
P(v_n) & \propto & v_n \exp \left( -\beta v_n^2 \right) \\
P(v_t) & \propto & \exp \left( -\beta v_t^2 \right), \label{eq_stochastic2}
\end{eqnarray}
so that the colloid acts as an additional thermostat~\cite{Lebo78}.
Such stochastic boundary conditions have been used for colloidal
particles by Inoue {\em et al.}~\cite{Inou02} and Hecht {\em et
al.}~\cite{Hech05}.  We have systematically studied several
implementations of stick boundary conditions for spherical
colloids~\cite{Padd05}, and derived a version of the stochastic
boundary conditions which reproduces linear and angular velocity
correlation functions that agree with Enskog theory for short times,
and hydrodynamic mode-coupling theory for long times.  We argue that
the stochastic rule of Eq.~(\ref{eq_stochastic1}) is more like a real
physical colloid -- where fluid-surface interactions are mediated by
steric stabilizing layers or local co- and counter-ion concentrations
-- than bounce-back rules are \cite{Padd05}.

Nevertheless, in this paper, many examples will be for radial
interactions such as those described in Eq.~(\ref{eq:col-fluid}).
These do not transfer angular momentum to a spherical colloid, and so
induce effective {\em slip boundary conditions}.  For many of the
hydrodynamic effects we will discuss here the difference
with stick boundary conditions is quantitative, not qualitative, and
also well understood.

\subsection{Depletion and lubrication forces}

\subsubsection{Spurious depletion forces induced by the fluid}

We would like to issue a warning that the additional fluid degrees of
freedom may inadvertently introduce depletion forces between the
colloids.  Because the number density of SRD particles is much
higher than that of the colloids, even a small overlap between two
colloids can lead to enormous attractions.

\begin{figure}[t]
  \scalebox{0.50}{\includegraphics{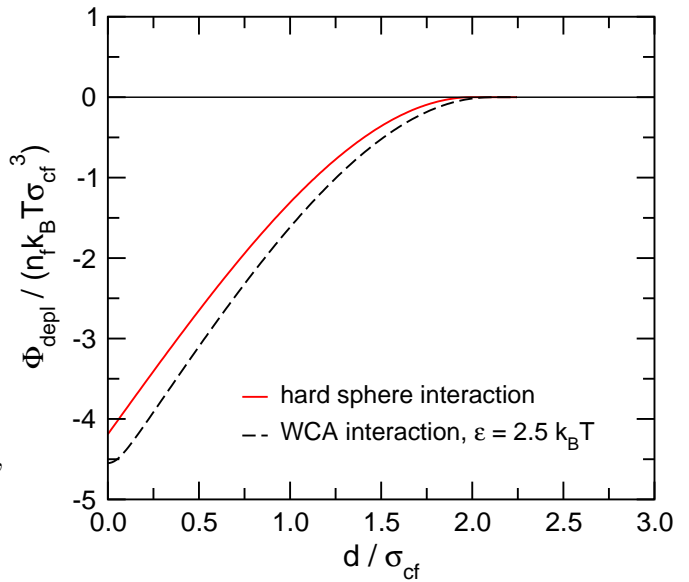}}
\caption{(Color online) Effective depletion potentials  induced between two colloids
by the SRD fluid particles, for HS fluid-colloid (solid
line) and WCA fluid-colloid (dashed line) interactions taken from
Eq.~(\protect\ref{eq:col-fluid}).  The interparticle distance is
measured in units of $\sigma_{cf}$. Whether these attractive potentials have
important effects or not depends on the choice of diameter $\sigma_{cc}$ in the bare colloid-colloid potential~(\protect\ref{eq:col-col}).
\label{fig_depletion}
}
\end{figure}

 For low colloid
densities the equilibrium depletion interaction between any two
colloids caused by the presence of the ideal fluid particles is given
by ~\cite{Asak58,Loui02}:
\begin{equation}
\Phi_{\mathrm{depl}}(d) = n_f k_B T \left[ V_{\mathrm{excl}}(d) - V_{\mathrm{excl}}(\infty) \right]\mbox{,}
\label{eq_depletion}
\end{equation}
where $n_f = \gamma/a_0^3$ is the number density of fluid particles
and $V_{\mathrm{excl}}(d)$ is the (free) volume excluded to the fluid
by the presence of two colloids separated by a distance $d$.  The
latter is given by
\begin{equation}
V_{\mathrm{excl}}(d) = \int \mathrm{d}^3\mathbf{r} \left\{ 1- \exp \left[ 
   -\beta \varphi_{cf} \left(\mathbf{r}-\mathbf{r}_1 \right)
   -\beta \varphi_{cf} \left(\mathbf{r}-\mathbf{r}_2 \right) \right] \right\}\mbox{,}
   \label{eq_overlap}
\end{equation}
where $\left| \mathbf{r}_1 - \mathbf{r}_2 \right| = d$. An example is
given in Fig.~\ref{fig_depletion} where we have plotted the resulting
depletion potential for the colloid-solvent interaction~(\ref{eq:col-fluid}), with $\epsilon_{cc}= \epsilon_{cf} = 2.5 k_BT$ as routinely used
in our simulations, as well as the depletion interaction resulting
from a truly HS colloid-solvent interaction. The latter can
easily be calculated analytically, with the result
\begin{eqnarray}
\Phi_{\mathrm{depl}}^{\mathrm{HS}}(d) & = & - n_f k_B T \frac{4}{3}\pi \sigma_{cf}^3
\left(1 - \frac{3d}{4\sigma_{cf}} + \frac{d^3}{16 \sigma_{cf}^3} \right)\nonumber \\
& \, & \qquad \mbox{for $d < 2\sigma_{cf}$.}
\end{eqnarray}

 For the pure HS interactions, one could take $\sigma_{cc} \geq 2
 \sigma_{cf}$ and the depletion forces would have no effect.  But for
 interactions such as those used in
 Eqs.~(\ref{eq:col-col})~and~(\ref{eq:col-fluid}), the softer
 repulsions mean that inter-colloid distances less than $\sigma_{cc}$
 are regularly sampled. A more stringent criterion of $\sigma_{cc}$
 must  therefore be used to avoid spurious depletion effects.  As an
 example, we re-analyze the simulations of ref.~\cite{Lee01}, where a
 WCA form with $n=6$ was used for the fluid-colloid interactions, and
 a normal Lennard Jones $n=6$ potential with $\sigma_{cf} \geq \frac12
 \sigma_{cc}$ was used for the colloid-colloid interactions.  The authors 
 found differences between ``vacuum'' calculations without SRD
 particles, and a hybrid scheme coupling the colloids to an SRD
 solvent.  These were correctly attributed to ``solvent induced
 pressure'', which we quantify here as depletion interactions.  For
 one set of their parameters, $\sigma_{cf} = \frac12 \sigma_{cc}$, the
 effect is mainly to soften the repulsion, but for their other
 parameter set: $\sigma_{cf} = 0.65 \sigma_{cc}$, depletion
 attractions induce an effective attractive well-depth of over $40 k_B
 T$!

Since the depletion potentials can be calculated analytically, one
might try counteracting them by introducing a compensating repulsive
potential of the form: $\Phi_{\mathrm{comp}} = -
\Phi_{\mathrm{depl}}$ between the colloids. However, there are three
problems with this approach: Firstly, at higher colloid packing fractions,
three and higher order interactions may also need to be added, and
these are very difficult to calculate.  Secondly, the depletion
interactions are not instantaneous, and in fact only converge to their
equilibrium average algebraically in time~\cite{Vlie03}.  While this
is not a problem for equilibrium properties, it will introduce errors
for non-equilibrium properties.  Finally, when external fields drive
the colloid, small but persistent anisotropies in the solvent density
around a colloid may occur~\cite{Dzub03}. Although these density
variations are (and should be) small, the resulting variations in
depletion interactions can be large.

To avoid these problems, we routinely choose the colloid-fluid
interaction range $\sigma_{cf}$ slightly below half the colloid
diameter $\sigma_{cc}/2$. More precisely, we ensure that the
colloid-colloid interaction equals $2.5 k_B T$ at a
distance $d$ where the depletion interactions have become zero, i.e.,
at a distance of twice the colloid-solvent interaction cut-off
radius. Smaller distances will consequently be rare, and adequately
dealt with by the compensation potential.  This solution may be a more
realistic representation anyhow, since in practice for charge and even
for sterically stabilized colloids, the effective colloid-colloid
diameter $\sigma_{cc}$ is expected to be larger than twice the
effective colloid-fluid diameter $\sigma_{cf}$.  This is particularly
so for charged colloids at large Debye screening lengths.

\subsubsection{Lubrication forces depend on surface details}

When two surfaces approach one another, they must displace the fluid
between them, while if they move apart, fluid must flow into the space
freed up between the surfaces.  At very short inter-surface distances,
this results in so-called lubrication forces, which are repulsive for
colloids approaching each other, and attractive for colloids moving
apart~\cite{Russ89,Dhon96,Bren83}. These forces are expected to be
particularly important for driven dense colloidal suspensions; see
e.g.~\cite{Verm05} for a recent review. 

An additional advantage of our choice of diameters $\sigma_{ci}$ above
is that more fluid particles will fit in the space between two
colloids, and consequently the lubrication forces will be more
accurately represented.  It should be kept in mind that for colloids,
the exact nature of the short-range lubrication forces will depend on
physical details of the surface, such as its roughness, or presence of
a grafted polymeric stabilizing layer~\cite{Pota95,Nomm99,Melr04}.  For
perfectly smooth colloids, analytic limiting expressions for the
lubrication forces can be derived~\cite{Bren83}, showing a divergence
at short distances.  We have confirmed that SRD resolves these
lubrication forces down to surprisingly low interparticle
distances. 
But, at some point, this will break down, depending of
course on the choice of simulation parameters (such as
$\sigma_{cf}/a_0$, $\lambda$, and $\gamma$), as well as the details of
the particular type of colloidal particles that one wishes to
model~\cite{Pota95,Nomm99,Melr04}. An explicit analytic correction
could be applied to properly resolve these forces for very small
distances, as was recently implemented for Lattice
Boltzmann~\cite{Nguy02}.  However, in this paper, we will assume that
our choice of $\sigma_{cf}$ is small enough for SRD to sufficiently
resolve lubrication forces.  The lack of a complete divergence at very
short distances may be a better model of what happens for real
colloids anyway.  For dense suspensions under strong shear, explicit
lubrication force corrections as well as other short-ranged
colloid-colloid interactions arising from surface details such as
polymer coats will almost certainly need to be put in by hand, see
e.g. ref.~\cite{Melr04} for further discussion of this subtle problem.

\subsection{Test of static properties}

\begin{figure}[t]
  \scalebox{0.50}{\includegraphics{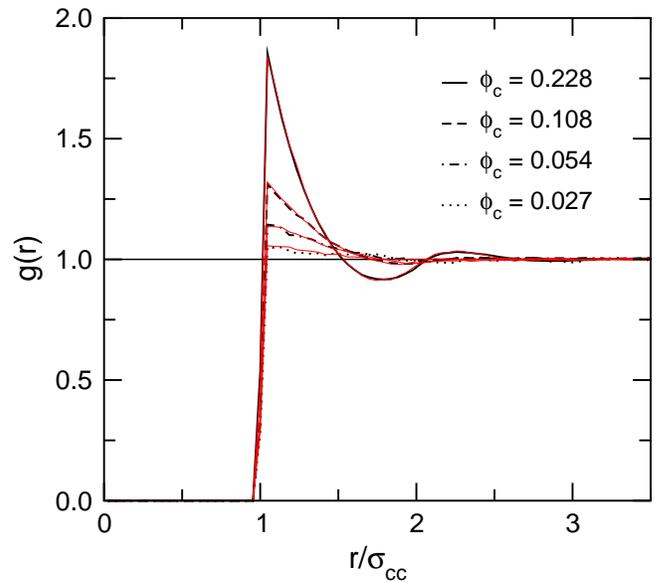}}
\caption{(Color online) Colloid radial distribution functions $g(r)$
 for various colloid volume fractions $\phi_c$.  The simulations were
 performed with BD (black lines) and SRD (red lines) simulations, and
 are virtually indistinguishable within statistical errors.
 \label{fig_gr} 
}
\end{figure}

The equilibrium properties of a statistical mechanical system should be
independent of the detailed dynamics by which it explores phase space.
Therefore one requisite condition imposed on our coarse-grained dynamics is
that it reproduces the correct static properties of an ensemble of colloids.

An obvious property to measure is the radial distribution function
$g(r)$~\cite{Hans86}.  In Fig.~\ref{fig_gr} we depict some $g(r)$s obtained from SRD
simulations at a number of different densities, and compare these to
similar simulations with standard Brownian dynamics. The colloids
interact via potentials of the
form~(\ref{eq:col-col})~--~(\ref{eq:col-fluid}) with $\sigma_{cf} =
2 a_0$, $\sigma_{cc} = 4.3 a_0$, and $\epsilon_{cc}=\epsilon_{cf} = 2.5 k_BT$. For the SRD
we used $\gamma =5, \alpha = \frac12 \pi$, $\Delta t_{\mathrm{MD}} = 0.025 t_0$ and $\Delta t_c = 0.1 t_0$,
implying $\lambda = 0.1$. For the BD we used a background friction calculated from the effective hydrodynamic
radius ({\it vide infra}) and a time-step equal to the $\Delta t_{\mathrm{MD}}$ used above. The colloids were placed in a
box of dimensions $32 \times 32 \times 96 a_0^3$, and the number of particles was varied from 64 to 540 in order to
achieve different packing fractions
\begin{equation}
\phi_c = \frac{1}{6}\pi n_c \sigma_{cc}^3,
\end{equation}
where the subscript in $\phi_c$ refers to volume fraction based on the
colloid-colloid interaction, to distinguished it from the volume
fraction $\phi$ based on the colloid hydrodynamic radius~\cite{BH}.
From Fig.~\ref{fig_gr} is clear that the two simulations are
indistinguishable within statistical errors, as required.  Even though
$\sigma_{cf} < \frac12 \sigma_{cc}$, we still found it necessary to
include an explicit compensating potential without which the $g(r)$
generated by SRD is increased noticeably at contact when compared to
BD simulations.

Finally, we note that any simulation will show finite size effects.
These may be thermodynamic as well as dynamic.  If there are
thermodynamic finite-size effects (for example due to a long
correlation length caused by proximity to a critical point), then a
correct simulation technique will show the same behaviour regardless
of the underlying dynamics.

\section{Dimensionless numbers}\label{Dimensionless}

Different regimes of hydrodynamic behavior can be characterized by a
series of dimensionless numbers that indicate the relative strengths of
competing physical processes~\cite{Hydrobook}.  If two distinct
physical systems can be described by the same set of hydrodynamic
numbers, then their flow behavior will be governed by the same
physics, even if their time and length scales differ by orders of
magnitude. In other words, a macroscopic boulder sedimenting in viscous molten
magma and a mesoscopic colloid sedimenting in water may show 
similar behavior if they share key hydrodynamic dimensionless
numbers.  It is therefore very instructive to analyze where a system
described by SRD sits in  ``hydrodynamic parameter space''.  In this
section we review this parameter space, one ``dimension'' at a time,
by exploring the hydrodynamic numbers listed in Table \ref{table:dimensionless}.

\begin{table}
\begin{center}
\caption{Dimensionless hydrodynamic numbers for colloidal suspensions.
\label{table:dimensionless}}
\begin{ruledtabular}
\begin{tabular}{lcll}
\noalign{\medskip}
Schmidt & $\displaystyle\frac{\mbox{collisional momentum
transport}}{\mbox{kinetic momentum transport}}$ & Sc $\displaystyle
=\frac{\nu}{ D_f}$  & Eq.~(\ref{eq:Schmidt}) \\ \noalign{\bigskip}

 Mach & $\displaystyle\frac{\mbox{flow velocity}}{\mbox{sound
 velocity}}$ & Ma $\displaystyle= \frac{v_s}{c_s}$ &
 Eq.~(\ref{eq:Mach}) \\ \noalign{\bigskip}

Reynolds & $\displaystyle\frac{\mbox{inertial forces}}{\mbox{viscous
forces}}$  & Re $\displaystyle = \frac{ v_s a}{\nu}$ & 
Eq.~(\ref{eq:Reynolds}) \\ \noalign{\bigskip}

 Knudsen & $\displaystyle\frac{\mbox{mean free path  }}{\mbox{particle
 size}}$ & Kn $\displaystyle =\frac{\lambda_{\mathrm{free}}}{a}$ &
 Eq.~(\ref{eq:Knudsen}) \\ \noalign{\bigskip}

 Peclet &$\displaystyle \frac{\mbox{convective
 transport}}{\mbox{diffusive transport}}$ & Pe $\displaystyle
 =\frac{v_s a}{D_{\mathrm{\mathrm{col}}}}$ & Eq.~(\ref{eq:Peclet})\\
 \noalign{\medskip}

\end{tabular}
\end{ruledtabular}
\end{center}
\end{table}

\subsection{Schmidt number}

The Schmidt number
\begin{equation}\label{eq:Schmidt}
\mbox{Sc} = \frac{\nu}{D_f}
\end{equation}
is important for characterizing a pure fluid.  It expresses the rate
of diffusive momentum transfer, measured by the kinematic viscosity $\nu$,
relative to the rate of diffusive mass transfer, measured by the fluid particle
self-diffusion coefficient $D_f$.  For a gas, momentum transport is
dominated by mass diffusion, so that Sc
$\approx 1$, whereas for a liquid, momentum transport is dominated by
inter-particle collisions, and Sc $\gg 1$.  For low Schmidt numbers,
the dynamics of SRD is indeed ``gas-like'', while for larger Sc, the
dynamics shows collective behavior reminiscent of the hydrodynamics of
a liquid~\cite{Ripo05}.  This distinction will be particularly
important for simulating the dynamics of embedded particles that
couple to the solvent through participation in the collision step.
For particles that couple directly, either through a potential, or
through bounce-back or stochastic boundary conditions, this
distinction is less important because the fluid-particle diffusion
coefficient $D_f$ does not directly enter the Navier Stokes equations.
Of course other physical properties can be affected. For example, we
expect that  the self diffusion coefficient of a colloid
$D_{\mathrm{col}}$ should be much smaller than the fluid diffusion
coefficient, i.e.\ $D_{\mathrm{col}}
\ll D_f$ to achieve the correct time-scale separation, as we will see in
Section~\ref{time-scales}.

\subsubsection{Dependence of Schmidt number on simulation parameters}

From Eqs.~(\ref{eq:diffusion})~--~(\ref{eq:fcol})
 it follows that
the Schmidt number   can be rewritten as
\begin{equation}\label{eq:Schmidt-scaling}
\mbox{Sc} \approx \frac{1}{3}  + \frac{1}{18 \lambda^2},
\end{equation}
where we have ignored the dependence on $\gamma$ and $\alpha$ 
since  the dominant scaling is with the dimensionless
mean-free path $\lambda$.  For larger $\lambda$, where the kinetic
contributions to the viscosity dominate, the Sc number is small, and
the dynamics is gas-like.  Because both the fluid diffusion
coefficient~(\ref{eq:diffusion}) and the kinetic contribution to the
viscosity~(\ref{eq:nukin}) scale linearly with $\lambda$, the only
way to obtain a large Sc number is in the limit $\lambda \ll 1$ where 
the collisional contribution to $\nu$ dominates.

Low Sc numbers may be a more general characteristic of particle-based
coarse-graining methods.  For computational reasons, the number of
particles is normally greatly reduced compared to the solvent one is
modeling. Thus the average inter-particle separation and mean-free
path are substantially increased, typically leading to a smaller
kinematic viscosity and a larger fluid diffusion coefficient $D_f$.
With DPD, for example, one typically finds $\mbox{Sc} \approx
1$~\cite{Dunw04,Lowe99}.  It is therefore difficult to achieve large
Sc numbers, as in a real liquid, without sacrificing computational
efficiency.  But this may not always be necessary. As long as momentum
transport is clearly faster than mass transport, the solvent should
behave in a liquid-like fashion.  With this argument in mind, and also
because the Sc number does not directly enter the Navier Stokes
equations we want to solve, we use $\lambda = 0.1$ in this paper and
in ref.~\cite{Padd04}, which leads to a Sc $\approx 5$ for $\gamma = 5$
and $\alpha = \pi/2$.  This should be sufficient for the problems we
study.  For other systems, such as those where the suspended particles
are coupled to the solvent through the collision step, more care may
be needed to ensure that the Sc number is indeed large enough \cite{Ripo05}.

\subsection{Mach number}

The Mach number measures the ratio
\begin{equation}\label{eq:Mach}
\mbox{Ma} = \frac{v_s}{c_f},
\end{equation}
between $v_s$, the speed of solvent or colloid flow, and
$c_f=\sqrt{(5/3) (k_B T/m_f)}$, the speed of sound.  In contrast to
the Schmidt number, which is an intrinsic property of the solvent, it
depends directly on flow velocity.  The Ma number measures
compressibility effects~\cite{Hydrobook} since sound speed is related
to the compressibility of a liquid.  Because $c_f$ in many liquids is
of order $10^3$ m/s, the Ma numbers for physical colloidal systems are
extremely small under normally achievable flow conditions.  Just as
for the Sc number, however, particle based coarse-graining schemes
drastically lower the Ma number. The particle mass $m_f$ is typically
much greater than the mass of a molecule of the underlying fluid, resulting in lower
velocities, and
moreover, due to the lower density, collisions also occur less
frequently. These effects mean that the speed of sound is much lower
in a coarse-grained system than it is in the underlying physical
fluid. Or, in other words, particle based coarse-graining systems
 are typically much more compressible than the solvents
they model.

Ma number effects typically scale with $\mbox{Ma}^2$~\cite{Hydrobook},
and so the Ma number does not need to be nearly as small as for a
realistic fluid to still be in the correct regime of hydrodynamic
parameter space.  This is convenient, because   to lower 
the Ma number, one would need to integrate over 
longer fluid particle trajectories  to allow, for example, a
colloidal particle to flow over a given distance, making
the simulation  computationally more expensive.  So there
is a compromise between small Ma numbers and computational efficiency.
We limit our Ma numbers to values such that $\mbox{Ma} \leq 0.1$, but it
might be possible, in some situations, to double or triple that limit
without causing undue error.  For example, incompressible hydrodynamics 
is used for aerodynamic flows up to such Ma numbers  since the errors
are expected to  scale as $1/(1-\mbox{Ma}^2)$~\cite{Hydrobook}.
When working in units of $m_f$ and $k_B T$, the only way to keep the
Ma number below our upper limit is to restrict the maximum flow velocity to
$v_s \lesssim 0.1 c_f$.  The flow velocity itself is, of course,
determined by the external fields, but also by  other
parameters of the system.

\subsection{Reynolds number}

The Reynolds number is one of the most important dimensionless numbers
characterizing hydrodynamic flows.  Mathematically, it measures the
relative importance of the non-linear terms in the Navier-Stokes
equations~\cite{Hydrobook}.  Physically, it determines the relative
importance of inertial over viscous forces, and can be expressed as:
\begin{equation}\label{eq:Reynolds}
\mbox{Re} = \frac{\ v_s a}{\nu}
\end{equation}
where $a$ is a length-scale, in our case, the hydrodynamic radius of a
colloid, i.e.\ $a \approx \sigma_{cf}$, and $v_s$ is a flow velocity.

For a spherical particle in a flow, the following heuristic argument
helps clarify the physics behind the Reynolds number: If the {\em Stokes time} 
\begin{equation}\label{eq:tauS}
t_S = \frac{\sigma_{cf}}{v_s},
\end{equation}
 it
takes a particle to advect over its own radius 
is about the same as the {\em kinematic time}
\begin{equation}\label{eq:taunu}
\tau_\nu=\frac{\sigma_{cf}^2}{\nu} = \mathrm{Re} \,\, t_S
\end{equation}
 it takes momentum to diffuse over that distance, i.e.\
 Re=$\tau_\nu/t_S \approx 1$, then the particle will feel vorticity
 effects from its own motion a distance $\sigma_{cf}$ away, leading to
 non-linear inertial contributions to its motion.  Since hydrodynamic
 interactions can decay as slowly as $1/r$, their influence can be
 non-negligible.  If, on the other hand, Re $\ll 1$, then vorticity
 will have diffused away and the particle will only feel very weak
 hydrodynamic effects from its own motion.

Exactly when inertial finite Re effects become significant depends
on the physical system under investigation.  For example, in a pipe,
where the length-scale $a$ in Eq.~(\ref{eq:Reynolds}) is its diameter, the
transition from simpler laminar to more complex turbulent flow is at a
pipe Reynolds number of Re $ \approx 2000$~\cite{Hydrobook,Mata03}.  On the
other hand, for a single spherical particle,
a non-linear dependence of the friction $\xi$ on the velocity
$v_s$, induced by inertial effects, starts to become noticeable for a particle Reynolds number of Re
$\approx 1$~\cite{Hydrobook}, while deviations in   the
 symmetry of the streamlines around a rotating sphere have been
 observed in calculations for Re $ \approx 0.1$~\cite{Miku04}.

For typical colloidal suspensions, where the particle diameter is on
the order of a few $\mu$m down to a few nm, the particle Reynolds
number is rarely more than $10^{-3}$.  In this so-called {\em Stokes
regime} viscous forces dominate and inertial effects can be completely
ignored. The Navier-Stokes equations can be replaced by the linear
Stokes equations~\cite{Hydrobook} so that analytic solutions are
easier to obtain~\cite{Bren83}. However, some of the resulting
behavior is non-intuitive for those used to hydrodynamic effects on a
macroscopic scale. For example, as famously explained by Purcell in a
talk entitled ``Life at low Reynolds numbers''~\cite{Purc}, many
simple processes in biology occur on small length scales, well into
the Stokes regime.  The conditions that bacteria, typically a few $\mu$m
long, experience in water are more akin to those humans would
experience in extremely thick molasses. Similarly, colloids, polymers,
vesicles and other small suspended objects are all subject to the same
physics, their motion dominated by viscous forces.  For example, if
for a colloid sedimenting at 1 $\mu$m/s, gravity were instantaneously
turned off, then the Stokes equations suggest that it would come to a
complete halt in a distance significantly less than one \AA, reflecting
the irrelevance of inertial forces in this low Re number regime.
It should be kept in mind that when the Stokes regime is reached
because of small length scales (as opposed to very large viscosities
such as those found for volcanic lava flows), then thermal
fluctuations are also important. These will drive diffusive
behavior~\cite{Berg}.  In many ways SRD is ideally suited for this
regime of small particles  because the
thermal fluctuations are naturally included~\cite{Male00}.

\subsubsection{Dependence of the Reynolds number on simulation parameters}
From Eqs.~(\ref{eq:Mach})~and~(\ref{eq:Reynolds}), it follows that the Reynolds number for a
colloid of hydrodynamic radius $a\approx \sigma_{cf}$ can be written as:
\begin{eqnarray}\label{eq:Re-scaling}
\mbox{Re} &=& \sqrt{\frac{5}{3}}\,\,\mbox{Ma}\,\left( \frac{\sigma_{cf}}{a_0}\right) \,
\left(\frac{\nu_0}{\nu}\right).  
\end{eqnarray}
 Equating the hydrodynamic radius $a$ to $\sigma_{cf}$ from the
fluid-colloid WCA interaction of Eq.~(\ref{eq:col-fluid}) is not quite
correct, as we will see in section~\ref{finite-size}, but is a good
enough approximation in light of the fact that these hydrodynamic
numbers are qualitative rather than quantitative indicators.

In order to keep the Reynolds number low, one must either use small
particles, or very viscous fluids, or low velocities $v_s$.  The
latter condition is commensurate with a low Ma number, which is also
desirable.  For small enough $\lambda$ (and ignoring for simplicty the 
factor 
$f_{\mathrm{col}}^\nu(\gamma,\alpha)$ in Eq.~(\ref{eq:nucol})) the Re
number then scales as
\begin{equation}\label{eq:Re-small-lambda}
\mbox{Re} \approx  23 \,\, \mbox{Ma}\,\, \frac{\sigma_{cf}}{a_0} \,\, \lambda.
\end{equation}
 Again, we see that a smaller mean-free path $\lambda$,
which enhances the collisional viscosity, also helps bring the Re
number down. 
This parameter choice is also consistent with a
larger Sc number.  Thus larger Sc and smaller Ma numbers both help
keep the Re number low for SRD.
  In principle a large  viscosity can
also be obtained for large $\lambda$, which enhances the kinetic
viscosity, but this choice also lowers the Sc number and raises the
Knudsen number, which, as we will see in the next sub-section, is not
desirable.

Just as was found for the Ma number, it is relatively speaking more
expensive 
computationally  to simulate for low Re numbers because the
flow velocity must be kept low, which means longer simulation times
are necessary to reach time-scales where the suspended particles or
fluid flows have moved a significant distance.  We therefore
compromise and normally keep Re $\lesssim 0.2$, which is similar to
the choice made for LB simulations~\cite{Ladd96,Cate04}.  For many
situations related to the flow of colloids, this should be
sufficiently stringent.

\subsection{Knudsen number}

The Knudsen number Kn measures rarefaction effects, and can be written as
\begin{equation}\label{eq:Knudsen}
\mbox{Kn} = \frac{\lambda_{\mathrm{free} }}{a}
\end{equation}
where $a$ is a characteristic length-scale of the fluid flow, and
$\lambda_\mathrm{free}$ is the mean-free path of the fluid or gas. For
a colloid in SRD one could take $a\approx\sigma_{cf}$ and
$\lambda_\mathrm{free} \approx \lambda a_0$.  For large Knudsen
numbers $\mbox{Kn} \gtrsim 10$ continuum Navier-Stokes equations
completely break down, but even for much smaller Knudsen numbers,
significant rarefaction effects are seen.  For example, for flow in a
pipe, where $a$ in Eq.~(\ref{eq:Knudsen}) is taken to be the pipe
radius, important non-continuum effects are seen when $\mbox{Kn}
\gtrsim 0.1$.  In fact it is exactly for these conditions of modest Kn
numbers, when corrections to the Navier Stokes become noticeable, that
DSMC approaches are often used~\cite{Bird70,Bird94}.  SRD, which is
related to DSMC, may also be expected to work well for such flows.  It
could find important applications in microfluidics and other
micromechanical devices where Kn effects are expected to play a
role~\cite{Zhan05,Squi05}.

Colloidal dispersions normally have very small Kn numbers because the
mean-free path of most liquid solvents is very small.  For water at
standard temperature and pressure $\lambda_{\mathrm{free}} \approx 
3$~\AA. Just as found for the other dimensionless numbers,
coarse-graining typically leads to larger Kn numbers because of the
increase of the mean-free path.  Making the Kn number smaller also
typically increases the computational cost because a larger number of
collisions need to be calculated.  In our simulations, we keep Kn$
\leq 0.05$ for spheres.  This rough criterion is based on the
observation that for small Kn numbers the friction coefficient on a
sphere is expected to be decreased by a factor $1 - \alpha$ Kn, where
$\alpha$ is a material dependent constant of order $1$~\cite{Bren83},
so that we expect Kn number effects to be of the same order as other
coarse-graining errors.  There are two ways to achieve small Kn
numbers: one is  by increasing $\sigma_{cf}/a_0$, the other is by
decreasing $\lambda$.  The second condition is commensurate with a
large Sc number or a small Re number.

In ref.~\cite{Hech05}, the Kn numbers for colloids were $\mbox{Kn}
\approx 0.5$ and $\mbox{Kn} \approx 0.8$,  depending on their
colloid-solvent coupling. Such large Kn numbers should have a
significant effect on particle friction coefficients, and this may
explain why these authors find that  volume fraction $\phi$ has less
of an effect on the sedimentation velocity than we do~\cite{Padd04}.

\subsection{Peclet number}

The Peclet number Pe measures the relative strength of convective
transport to diffusive transport.  For example, for a colloid of
radius $a$, travelling at an average velocity $v_s$, the Pe
number is defined as:
\begin{equation}\label{eq:Peclet}
\mbox{Pe} = \frac{v_s a }{D_{\mathrm{col}}}
\end{equation}
where $D_{\mathrm{col}}$ is the colloid diffusion coefficient.
  Just as for the  Re number,
the Pe number can be interpreted as a ratio of a diffusive to a
convective time-scale, but now the former time-scale is not for the diffusion
 of momentum but rather it is given by the {\em colloid diffusion time}
\begin{equation}\label{eq:tauD}
\tau_D=\frac{\sigma_{cf}^2}{D_{\mathrm{col}}}
\end{equation}
which measures how long it takes for a colloid to diffuse over a
distance $\sigma_{cf}$. We again  take $a \approx \sigma_{cf}$ so
that, using Eq.~(\ref{eq:tauS}), the Pe number can be written as:
\begin{equation}\label{eq:Petime}
\mbox{Pe} = \frac{\tau_D}{t_S}.
\end{equation}
If Pe $\gg 1$ then the colloid moves convectively over a distance much
larger than its radius $\sigma_{cf}$ in the time $\tau_D$ that it
diffuses over that same distance.  Brownian fluctuations are expected
to be less important in this regime.  For Pe $\ll 1$, on the other hand,
the opposite is the case, and the main transport mechanism is
diffusive (note that on long enough time-scales ($t > \tau_D/\mathrm{Pe}^2$)
convection will always eventually ``outrun'' diffusion~\cite{Berg}).
It is sometimes thought that for low Pe numbers hydrodynamic effects
can be safely ignored, but this is not always true.  For example, we
found that the reduction of average sedimentation velocity with
particle volume fraction, famously first explained by
Batchelor~\cite{Batc72}, is independent of Pe number down to Pe $=0.1$
at least~\cite{Padd04}.

\subsubsection{Dependence of Peclet number on simulation parameters}

The highest Pe number achievable in simulation is limited by the constraints
on the Ma and Re numbers.  For example the Ma number sets an upper limit on
the maximum Pe number by limiting $v_s$.   From
 Eqs.~(\ref{eq:Reynolds})~and~(\ref{eq:Peclet}), it follows that the 
Peclet number can be re-written in terms of the Reynolds number as
\begin{equation}\label{eq:Reynolds-Peclet}
\mbox{Pe} = 
\frac{ \nu}{D_{\mathrm{col}}} \mathrm{Re} \approx 
 6 \pi \gamma \left( \frac{\nu}{\nu_0}\right)^2 \left(\frac{\sigma_{cf}}{a_0}\right)  \mbox{Re} 
\end{equation}
where we have approximated $D_{\mathrm{col}} \approx k_B T/(6 \pi \eta
\sigma_{cf})$.  This shows that for a given constraint on the Re number,
increasing $\gamma$ or $\sigma_{cf}$ increases the range of accessible
Pe numbers.  Similarly, when the kinematic viscosity is dominated by
the collisional contribution, decreasing the dimensionless mean-free
path $\lambda$ will also increase the maximum Pe number allowed since
$\mbox{Pe}^{max}\sim \lambda^{-2} \mbox{Re} \sim \lambda^{-1}\mbox{Ma}$.  However,
these changes increase the computational cost of the simulation.

\section{finite-size, discretization, and inertial effects}\label{finite-size}

The cost of an SRD simulation scales almost linearly with the number of
fluid particles $N_f$ in the system, and this contribution is usually
much larger than the cost of including the colloidal degrees of
freedom.  To optimize the efficiency of a simulation, one would
therefore like to keep $N_f$ as small as possible. This objective can
be achieved by keeping $\sigma_{cf}/a_0$ and the box-size $L$ small.
Unfortunately both these choices are constrained by the errors they
introduce.  Reducing $\sigma_{cf}/a_0$ means that short-ranged
hydrodynamic fields become less accurately resolved due to Kn number and
{\em discretization effects}.  Decreasing the box-size $L$ falls
foul of the long-ranged nature of the HI, and therefore, just as for
Coulomb interactions~\cite{AllenTildesley}, {\em finite size effects}
such as those induced by periodic images must be treated with special
care.

Increasing the flow velocity may also be desirable since more Stokes
times $\tau_S = \sigma_{cf}/v_s$ can be achieved for the same number of SRD
collision steps, thus  increasing computational efficiency. The
Ma number gives one constraint on $v_s$, but usually the more
stringent constraint comes from keeping the Re number low to prevent
unwanted {\em inertial effects}.

\begin{figure}[t]
  \scalebox{0.50}{\includegraphics{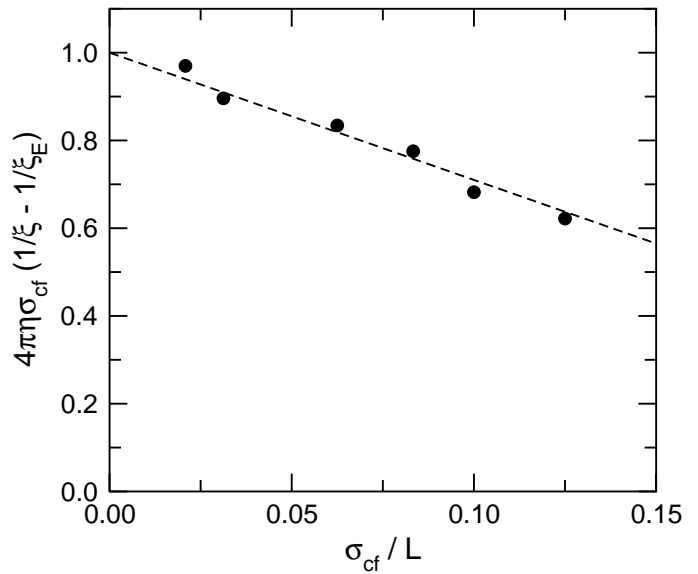}}
\caption{Finite system size scaling of the Stokes friction $\xi_S$ on a sphere of size
$\sigma_{cf} = 2$ in a flow of velocity $v_s = 0.01\,a_0/t_0$.  The
correction factor $f$ defined in Eq.~(\protect\ref{eq_xi_S}) can
be extracted from measured friction $\xi$ (drag force/velocity) and
estimated Enskog friction $\xi_E$ defined in Eq.~(\protect\ref{eq_xi_E}). It
compares well to theory, Eq.~(\protect\ref{eq:f}).
\label{fig_stokesboxscaling}
}
\end{figure}

\subsection{Finite-size effects}

\subsubsection{Finite-size correction to the friction}

The friction coefficient $\xi$ can be extracted from the Stokes drag
$\mathbf{F}_d$ on a fixed colloid in fluid flow:
\begin{equation}
\mathbf{F}_d = -\xi \mathbf{v}_{\infty} \equiv -4 \pi\eta a \mathbf{v}_{\infty}
\label{eq_F}
\end{equation}
where $\mathbf{v}_{\infty}$ is the flow field at large distances.  The
pre-factor 4 comes from using slip boundary conditions; it would be 6
for stick-boundary conditions, as verified in~\cite{Padd05}.  In principle, since $\eta$ is accurately known from
theory for SRD, this expression can be used to extract the
hydrodynamic radius $a$, which is not necessarily the same as
$\sigma_{cf}$, from a simulation, as we did in~\cite{Padd04}.

The hydrodynamic radius $a$ can also be directly calculated from
theory.  To derive this, it is important to recognize that there are
two sources of friction~\cite{Male00}.  The first comes from the local
Brownian collisions with the small particles, and can be calculated by
a simplified Enskog/Boltzmann type kinetic theory~\cite{Hyne77}:
\begin{equation}
\xi_E  =  \frac{8}{3} \left( \frac{2 \pi k_BT M_c m_f}{M+m_f} \right)^{1/2} n_f \sigma_{cf}^2,
\label{eq_xi_E}
\end{equation}
which is here adapted for slip boundary conditions.  A related expression 
for stick boundary conditions,
including that for rotational frictions, is described in~\cite{Padd05},
where it was shown that the short-time exponential decays of the 
linear  and angular  velocity auto-correlation functions are quantitatively described
by Enskog theory.

The second contribution to the friction, $\xi_S$, comes from
integrating the Stokes solution to the hydrodynamic field over the
surface of the particle, defined here as $r=\sigma_{cf}$.  These two
contributions to the friction should be added in parallel to obtain
the total friction~\cite{Hyne77,Lee04} (see also Appendix B):
\begin{equation}
\frac{1}{\xi} = \frac{1}{\xi_S} + \frac{1}{\xi_E} \label{eq_xi} .
\end{equation}

In contrast to the Enskog friction, which is local, we expect
substantial box-size effects on the Stokes friction $\xi_S$ since it
depends on long-ranged hydrodynamic effects.  
These can be expressed in terms of a correction factor
 $f(\sigma_{cf}/L)$ that should go to 1
 for very large systems :
\begin{equation}
\xi_S  =  4\pi \eta \sigma_{cf} f^{-1}(\sigma_{cf}/L) \label{eq_xi_S}.
\end{equation}
 To measure this correction factor we plot, in
 Fig.\ref{fig_stokesboxscaling}, the form $4\pi \eta \sigma_{cf}
 \left(1/\xi - 1/\xi_E\right)$ for various system box sizes for which
 we have measured $\xi$ from the simulation, and estimated $\xi_E$
 from Eq.~(\ref{eq_xi_E}). As expected, the correction factor tends to 1
 for smaller $\sigma_{cf}/L$.  More detailed
 calculations~\cite{Zick82,Duen93}, taking into account the effect of
 periodic boundaries, suggests that to lowest order in $\sigma_{cf}/L$
 the correction factor should scale as
\begin{equation}\label{eq:f}
f(\sigma_{cf}/L) \approx 1 + 2.837 \frac{\sigma_{cf}}{L}.
\end{equation}
 Indeed, a least squares fit of the data to this form gives a slope of
 $c \approx 2.9$, close to the theoretical value and  in agreement with
 similar Lattice Boltzmann simulations of a single colloidal
 sphere~\cite{Loba04}. 

\begin{figure}[t]
  \scalebox{0.50}{\includegraphics{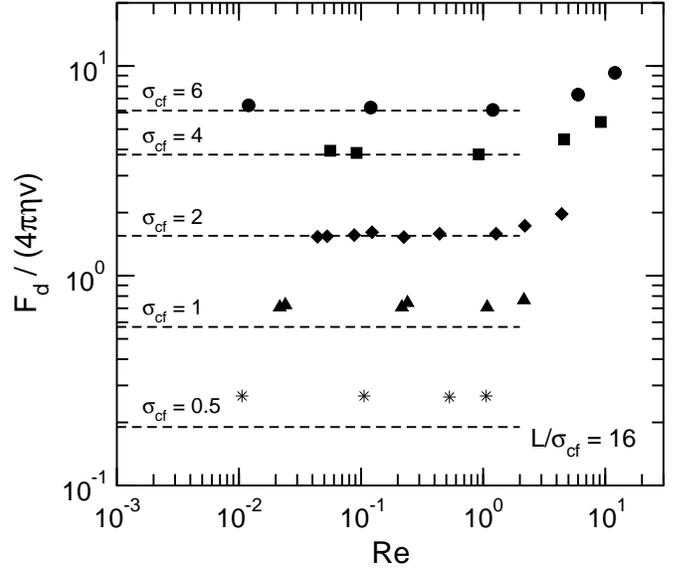}}
\caption{Drag force $F_d$ divided by $4\pi \eta v$ for various colloid sizes and
Reynolds numbers (in all cases the box size $L = 16 \sigma_{cf}$).
For small Reynolds numbers this ratio converges to the
effective hydrodynamic radius $a$.  Dashed lines are theoretical predictions
from combining Stokes and Enskog frictions in~(\protect\ref{eq_xi}).
\label{fig_hydroradius}
}
\end{figure}

With these ingredients in hand, we can calculate the theoretical
expected friction from Eqns.~(\ref{eq_xi_E})~-~(\ref{eq:f}).  We know
from previous work that the Enskog contribution at short times and the
hydrodynamic contribution at long times quantitatively reproduce the
rotational and translational velocity auto-correlation functions
(VACF) for $\sigma_{cf}/a_0=2-5$~\cite{Padd05}, and so we expect that
the friction coefficients should also be accurately described by these
theories.  We test this further in Fig.~\ref{fig_hydroradius} for a
number of different values of $\sigma_{cf}/a_0$, and find excellent
agreement with theory  for $\mathrm{Re} \leq 1$ and
$\sigma_{cf}/a_0 \geq 2$. For $\sigma_{cf}/a_0=1$ and below, on the
other hand, we find deviations from the theory.  These are most likely
due to Kn number and discretization effects, to be discussed in the
next sub-section.

For the smallest spheres the mass ratio $M_c/m_f$ may also change the
measured friction if instead of fixing the sphere we were to let it
move freely.  For $\sigma_{cf}= a_0$, $M_c/m_f \approx 21$, which is
small enough to have a significant effect~\cite{Schm03}.  For larger
spheres, this is not expected to be a problem.  For example $M_c/m_f
\approx 168 $ for $\sigma_{cf} = 2 a_0$ and $M_c/m_f
\approx 1340 $  for $\sigma_{cf} = 4 a_0$.

\subsubsection{Effective hydrodynamic radius}

From the calculated or measured friction we can also obtain the
effective hydrodynamic radius $a_{\mathrm{eff}}$, which is continuum concept, 
 from a comparison of the microscopic frictions from Eq.~(\ref{eq_xi})
with  Eq.~(\ref{eq_F}):
\begin{equation}\label{eq:effectiveradius}
a_{\mathrm{eff}} = \frac{\xi_S \xi_E}{4 \pi \eta \left(\xi_S + \xi_E\right)}
\end{equation}
(for stick-boundaries the $4 \pi$ should be replaced by $6 \pi$).
   We find
$a_{\mathrm{eff}}(\sigma_{cf}=6) = 6.11a_0$, $a_{\mathrm{eff}}(\sigma_{cf}=4)=3.78a_0$, and
$a_{\mathrm{eff}}(\sigma_{cf}=2)=1.55a_0$.  The effective hydrodynamic radius is
increased by the finite-size effects, but lowered by the Enskog
contribution which is added in parallel.  Because this latter
contribution is relatively more important for small colloids, $a_{\mathrm{eff}} <
\sigma_{cf}$ for smaller $\sigma_{cf}$.  For $\sigma_{cf}=6 a_0$ the
Enskog contribution is smaller than the finite-size effects, and so
the effective hydrodynamic radius is larger than $\sigma_{cf}$.
Obviously this latter effect depends on the box-size.  In an infinite
box $a_{\mathrm{eff}} < \sigma_{cf}$ for all values of $\sigma_{cf}$, due to 
the Enskog contribution.  Note
that it is the effective hydrodynamic radius $a_{\mathrm{eff}}$ which sets the
long-ranged hydrodynamic fields, as we will see below.

\subsubsection{Turning off long-ranged hydrodynamics}

Hydrodynamic forces can be turned off in SRD by regularly randomizing
the absolute fluid particle velocities (with for example a naive
Langevin thermostat). For particles embedded in an SRD solvent through
participation in the collision step, this trick can be used to compare
the effects of hydrodynamics to a purely Brownian
simulation\cite{Kiku03}.  The Yeomans group has successfully applied
this idea in a study of polymer collapse and protein
folding~\cite{Kiku05}.

  However for the case of the colloids embedded through direct solvent
  collisions, turning off the hydrodynamic forces by randomizing the
  velocities greatly enhances the friction because, as is clear from
  Eqns.~(\ref{eq_xi_E})~-~(\ref{eq_xi_S}), the two contributions add
  in parallel.  Without long-ranged hydrodynamics, the friction would
  be entirely dominated by the Enskog contribution~(\ref{eq_xi_E})
  which scales with $\sigma_{cf}^2$, and can be much larger than the
  hydrodynamic contribution which scales as $\sigma_{cf}$.  Another
  way of stating this would be: By locally conserving momentum, SRD
  allows the development of long-ranged hydrodynamic fluid velocity
  correlations that greatly reduce the friction felt by a larger
  colloidal particle compared to the friction it would feel from a
  purely random Brownian heat-bath at the same temperature and number
  density $n_f=\gamma/a_0^3$.

\subsection{Discretization effects on flow-field and friction}

\begin{figure}[t]
  \scalebox{0.50}{\includegraphics{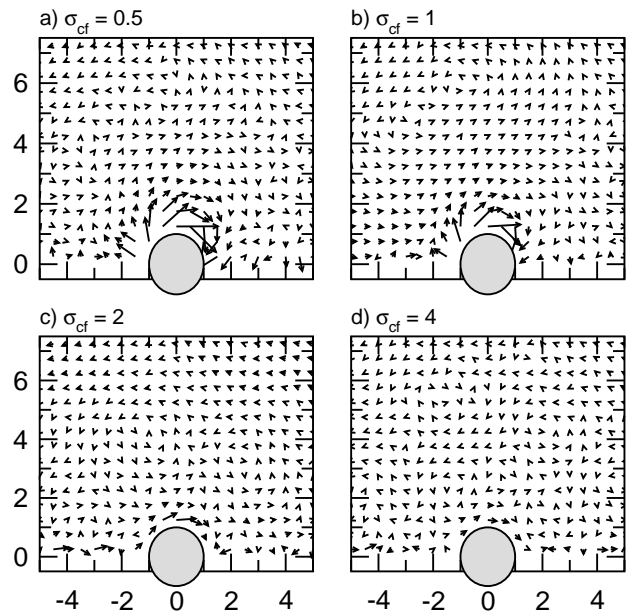}}
\caption{Magnified difference fields
$10 [\mathbf{v}(\mathbf{r}) - \mathbf{v}_{\mathrm{St}}(\mathbf{r})] / v_{\infty}$
for different colloidal sizes (the axes are scaled by $\sigma_{cf}$).
In all cases Re $\approx 0.1$, and box size $L = 16 \sigma_{cf}$. 
\label{fig_flowfield_a}
}
\end{figure}

  For pure Stokes flow (Re = 0), the velocity around a fixed
  slip-boundary sphere of (effective) hydrodynamic radius $a$ can be exactly
  calculated:
\begin{equation}
\mathbf{v}_{\mathrm{St}}(\mathbf{r}) = \mathbf{v}_{\infty} \left( 1 - \frac{a}{2r}\right) - 
\mathbf{v}_{\infty} \cdot \hat{\mathbf{r}} \hat{\mathbf{r}} \frac{a}{2r}
\label{eq_Stokes}
\end{equation}
where $\mathbf{v}_{\infty}$ is the velocity field far away from the
sphere, and $\mathbf{r}$ the vector pointing from the center of the sphere
to a position inside the fluid, with corresponding unit vector
$\hat{\mathbf{r}} = \mathbf{r}/r$.

In Fig.~\ref{fig_flowfield_a} we plot the difference between the
measured field and the theoretical expected field~(\ref{eq_Stokes})
for four different colloid radius to cell size ratios
$\sigma_{cf}/a_0$, using the values of the effective hydrodynamic
radius $a$ calculated from combining Eq.~(\ref{eq_xi}) and
Eq.~\ref{eq_F}.  As expected, the field is more accurately reproduced
as the colloid radius becomes larger with respect to the cell size $a_0$. This
improvement arises because SRD discretization and hydrodynamic Knudsen
number effects become smaller.  In Fig.~\ref{fig_flowfield_a} we
observe quite large deviations in the hydrodynamic field for
$\sigma_{cf}
\leq 1$.  These discretization effects 
may explain why the measured frictions in Fig.~\ref{fig_hydroradius}
do not agree with theory for these smallest sphere sizes.  We also
note that using $a=\sigma_{cf}$ instead of the more accurate value of
the effective hydrodynamic radius $a$ calculated from theory results
in significantly larger deviations between measured and theoretical
flow-fields.  This independently confirms the values of the effective
hydrodynamic radius.

The increased accuracy from using larger  $\sigma_{cf}/a_0$
comes at a sharp increase in computational
cost.  To make sure that the finite size effects in each simulation of
Fig.~\ref{fig_flowfield_a} are approximately the same, the box-size
was scaled as $L/\sigma_{cf}$. This means that doubling the colloid
size leads to an eightfold increase in the number of fluid
particles. Moreover, the maximum velocity of the fluid must go down
linearly in colloid size to keep the Re number, defined in
Eq.~(\ref{eq:Reynolds}), constant.  If in addition we keep the number
of Stokes times fixed, meaning that the fluid flows a certain multiple
of the colloid radius or the box size, then a larger particle also
means the fluid needs to flow over a proportionally longer total distance. The
overall computational costs for this calculation then scales at least
as $(\sigma_{cf}/a_0)^5$,  which is
quite steep.  For that reason, we advocate using smaller colloids
wherever possible.

In most of our simulations we choose $\sigma_{cf} = 2$, which leads to
a small relative error in the full velocity field and for which we can
fully explain the observed friction (as seen in the previous
subsection). This size is similar to what is commonly used in
LB~\cite{Ladd96,Cate04}.

\subsection{Inertial effects on flow-field and friction}

\begin{figure}[t]
  \scalebox{0.50}{\includegraphics{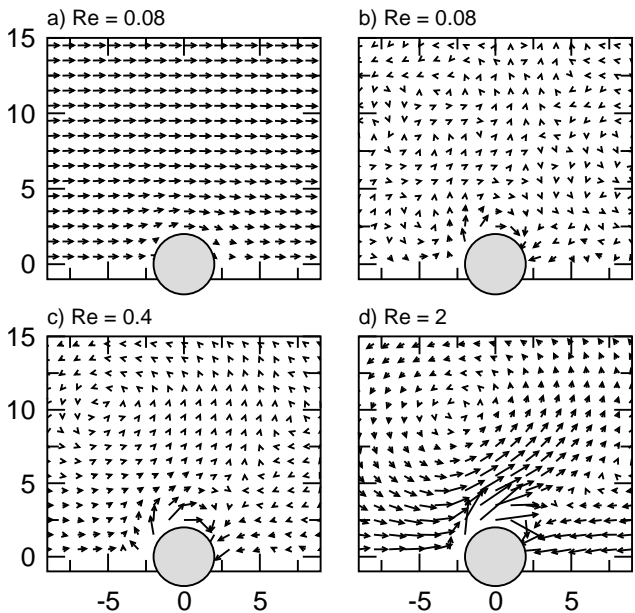}}
\caption{(a) Fluid flow-field around a fixed colloid with $\sigma_{cf}=2$ 
and $L=64 a_0$ for Re = 0.08.  (b)-(d) Difference field $10
[\mathbf{v}(\mathbf{r}) -
\mathbf{v}_{\mathrm{St}}(\mathbf{r})] / v_{\infty}$ for Re = 0.08,
0.4, and 2, respectively.  \label{fig_flowfield_Re} }
\end{figure}

One of the challenges in SRD is to keep the Re number down. It is
virtually impossible to reach the extremely low Re number (Stokes)
regime of realistic colloids, but that is not necessary either.  In
Fig.~\ref{fig_hydroradius} we see that the friction only begins to
noticeably vary from the Stokes limit at Re $\approx 1$ (at least on a
logarithmic scale).  We expect the flow-field itself to be more sensitive to
finite Re number effects.  To study this directly, we examine, in
Fig.~\ref{fig_flowfield_Re}, the flow-field around a fixed colloid of
size $\sigma_{cf} = 2 a_0$ in a box of $L=32a_0$ for for different
values of Re.  The differences with the Stokes flow field of
Eq.~(\ref{eq_Stokes}) are shown in Fig.~\ref{fig_flowfield_Re}(b)-(d)
for Re $=$ $0.08$, $0.4$, and $2$ respectively.  In all cases we used
a hydrodynamic radius of $a=1.55$ for the theoretical comparison, as
explained in the previous sub-section.  The lengths of the vectors in
Figs. (b)-(d) are multiplied by 10 for clarity.  We observe that the
relative errors increase with Re. They are on the order of $5\%$ for
Re=0.08, and increase to something on the order of $40\%$ for Re=2.
This is exactly what is expected, of course, as we are moving away
from the Stokes regime with which these flow lines are being compared.
Since we also expect effects on the order of a few \% from Kn number,
Ma number and various finite size effects, we argue that keeping
Re$\leq 0.2$ should be good enough for most of the applications we
have in mind.

\begin{table*}
\begin{center}
\caption{Time-scales  relevant for colloidal suspensions
\label{table:time}
}
\begin{ruledtabular}
\begin{tabular}{llll}

\begin{tabular}{c}
Solvent time-scales   
   \\ \hline \noalign{\medskip}
  \end{tabular}

 \\  

{\it  Solvent collision  time} over which solvent molecules 
interact & $\displaystyle \tau_{\mathrm{col}}  \approx 10^{-15}$~s & 
  \\ \noalign{\medskip} 

{\it  Solvent relaxation time} over which solvent velocity correlations
decay & $\displaystyle \tau_f \approx 10^{-14}- 10^{-13}$~s & \\  
\noalign{\medskip} \hline

 \begin{tabular}{c}
Hydrodynamic time-scales   
   \\ \hline \noalign{\medskip}
  \end{tabular}

 \\  

{\it Sonic time} over which sound propagates one colloidal radius &
$\displaystyle t_{cs} = \frac{a}{c_s} $ &
Eq.~(\protect\ref{eq:tcs})  \\  \noalign{\medskip}

{\it Kinematic time} over which momentum (vorticity) diffuses one colloidal radius  &
$\displaystyle \tau_{\nu} = \frac{a^2}{\nu} $ &
Eq.~(\protect\ref{eq:taunu})  \\  \noalign{\medskip}

  {\it Stokes time} over which a colloid convects over its own
  radius & $\displaystyle t_S = \frac{a}{v_s} = \frac{\tau_D}{\mbox{Pe}}$  & Eq.~(\protect\ref{eq:tauS})\\ 
  \noalign{\medskip}
\hline

\begin{tabular}{c}
Brownian time-scales   
   \\ \hline \noalign{\medskip}
  \end{tabular}
\\

{\it Fokker-Planck time} over which force-force correlations decay &
$\displaystyle \tau_{FP} $ & \\  \noalign{\medskip}

{\em Enskog relaxation time} over which short-time colloid velocity correlations decay  & $\displaystyle \tau_E = \frac{M_c}{\xi_E}$ & Eq.~(\protect\ref{eq:Enskog}) \\  \noalign{\medskip}

{\em Brownian relaxation time} over which colloid velocity correlations decay
in 
 Langevin Eq. & $\displaystyle \tau_B = \frac{M_c}{\xi_S}$ & Eq.~(\protect\ref{eq:tauB}) \\  \noalign{\medskip}

 {\it Colloid diffusion time} over which a colloid diffuses over its radius
  & $\displaystyle \tau_D = \frac{a^2}{D_{\mathrm{col}}}$ & Eq.~(\protect\ref{eq:tauD}) \\ \noalign{\medskip} \hline

 \begin{tabular}{c}
  Ordering of time-scales for colloidal particles  \\ \hline 
\end{tabular}  \\ \noalign{\medskip}

  $\displaystyle \tau_{\mathrm{col}} \,\,\,\,\, < \,\,\,\,\,  \tau_{f} , \tau_{FP}\,\,\,\,\ < \,\,\,\,\, \tau_{E},\,t_{cs} \,\,\,\,\, < \,\,\,\,\,  \tau_{B}  \,\,\,\,\, <  \,\,\,\,\, \tau_\nu  \,\,\,\,\,  <  \,\,\,\,\, \tau_D, \,\, t_S $ 

\end{tabular}
\end{ruledtabular}

\end{center}
\end{table*}

\section{The hierarchy of time scales}\label{time-scales}

\subsection{Colloidal time-scales}

Many different time-scales govern the physics of a colloid of mass $M$
embedded in a solvent~\cite{Dhon96}. The most important ones are
summarized in table~\ref{table:time}, and discussed in more detail below.

\subsubsection{Fluid time-scales}

 The shortest of these is  the  {\em solvent collision time}  $\tau_{\mathrm{col}}$ over
which fluid molecules interact with each other.  For a typical
molecular fluid, $\tau_{\mathrm{col}}$ is on the order of a few tens of fs, and any
MD scheme for a molecular fluid must use a discretization time-step $t_{MD} \ll
\tau_{\mathrm{col}}$ to properly integrate the equations of motion.
  
The next
time-scale up is the {\em solvent relaxation time} $\tau_f$, which measures
how fast the solvent VACF decays.  For a
typical molecular liquid, $\tau_f \approx 10^{-14} - 10^{-13}$ s.  (For water,
 at room temperature, for example, it is  on the order of 
50 femtoseconds) 

\subsubsection{Hydrodynamic time-scales for colloids}

Hydrodynamic interactions propagate by momentum (vorticity) diffusion, and 
also by sound.  The {\em sonic time} 
\begin{equation}\label{eq:tcs}
t_{cs} = \frac{a}{c_s}.
\end{equation}
it takes a sound wave to travel the radius of a colloid is typically
very small, on the order of 1 ns for a colloid of radius $a=1 \mu$m.
For that reason, sound effects are often ignored for colloidal
suspensions under the assumption that they will have dissipated so
quickly that they have no  noticeably influence on the dynamics.  However,
some experiments~\cite{Hend02} and theory~\cite{Bakk02} do find effects
from sound waves on colloidal hydrodynamics, so that this issue is not
completely settled yet.

The {\em kinematic time}, $\tau_\nu$, defined in Eq.~(\ref{eq:taunu})
(and Table~\ref{table:time})
as the time for momentum (vorticity) to diffuse over the radius of a
colloid, is particularly important for hydrodynamics.  It
sets the time-scale over which hydrodynamic interactions develop.
For a colloid of radius $1\mu$m, $\tau_\nu \approx 10^{-6}$ s,
which is much faster the colloidal diffusion time $\tau_D$.

When studying problems with a finite flow velocity, another
hydrodynamic time-scale emerges. The Stokes time $t_S$, defined in
Eq.~(\ref{eq:tauS}) (and Table~\ref{table:time}) as the time for a
colloid to advect over its own radius, can be related to the kinematic
time by the relation $\tau_\nu =
\mbox{Re}\,\, t_S$. Because colloidal particles are in the Stokes regime where
Re $\ll 1$, we find $\tau_\nu \ll t_S$.
 
Simulation and analytical methods based on the Oseen tensor and its
generalizations, e.g.~\cite{Erma78,Brad88}, implicitly assume that the
hydrodynamic interactions develop instantaneously, i.e.\ that
$\tau_\nu\approx 0$.  For the low Re number regime of colloidal
dispersions this is indeed a good approximation.  Of course it must be
kept in mind that $\tau_\nu$ is a diffusive time-scale, so that the
distance over which it propagates grows as $\sqrt{t}$. Thus the
approximation of instantaneous hydrodynamics must be interpreted with
some care for effects on larger length-scales.

\subsubsection{Brownian time-scales for colloids}

The fastest time-scale relevant to the colloidal Brownian motion is
the {\em Fokker-Planck time} $\tau_{FP}$ defined in~\cite{Dhon96} as
the time over which the force-force correlation function decays. It is
related to $\tau_f$, the time over which the fluid looses memory of
its velocity because the forces on
the colloid are caused by collisions with the fluid particles, 
 velocity differences de-correlate on a time scale
of order $\tau_f$.

 The next time-scale in this series is the {\em Brownian time}:
\begin{equation}\label{eq:tauB}
\tau_B = \frac{M_c}{\xi_S}
\end{equation}
where $\xi_S = 6 \pi \eta a$ is the Stokes friction for stick boundary
conditions.  It is often claimed that this measures the time for a
colloid to lose memory of its velocity.  However, as discussed in
Appendix~\ref{Langevin}, this picture, based on the Langevin equation,
is in fact incorrect for colloids. Nevertheless, $\tau_B$ has the
advantage that it can easily be calculated and so we will use it as a
crude upper bound on the colloid velocity de-correlation time.

Perhaps the most important time-scale relevant for Brownian motion is
the {\em diffusion time} $\tau_D$, described in Eq.~(\ref{eq:tauD}) as
the time for a colloidal particle to diffuse over its radius.  For a
colloid of radius $1 \mu$m, $\tau_D \approx 5$ s, and even though
$\tau_D \sim a^3$, it remains much larger than most microscopic
time-scales in the mesoscopic colloidal regime.

\subsection{Time-scales for coarse-grained simulation}

In the previous subsection we saw that the relevant time-scales for a
single colloid in a solvent can span as many as 15 orders of
magnitude.  Clearly it would be impossible to bridge all the
time-scales of a physical colloidal system -- from the molecular
$\tau_{\mathrm{col}}$ to the mesoscopic $\tau_D$ -- in a single
simulation.  Thankfully, it is not necessary to exactly reproduce each
of the different time-scales in order to achieve a correct
coarse-graining of colloidal dynamics. As long as they are clearly
separated, the correct physics should still emerge.

Since we take the view that the ``solvent'' particles are really a
Navier Stokes solver with noise, what is needed to reproduce Brownian
behavior is first of all that the colloid experiences random kicks
from the solvent on a short enough time-scale.
 By dramatically
reducing the number of ``molecules'' in the solvent, the number of
kicks per unit of time is similarly reduced.  Here we identify
the Fokker-Planck time $\tau_{FP}$ as the time-scale on which the
colloid experiences random Brownian motion.  For Brownian motion it
doesn't really matter how the kicks are produced, they could be completely
uncorrelated, but since we also require that the solvent
transports momentum, solvent particles must have some kind of
correlation, which in our case is represented by the time $\tau_f$.
Other colloid relaxation times should be much longer than this.  We
therefore require first of all that $\tau_{FP} \ll \tau_B$ and $\tau_f
\ll \tau_B$.

Secondly, the colloid's VACF should decay to zero well before it has
diffused or convected over its own radius.  A proper separation of
time-scales in a coarse-graining scheme then requires that $\tau_B \ll
\tau_D$ as well as $\tau_B \ll t_S$ for systems where convection is
important. 

\begin{figure}[t]
  \scalebox{0.4}{\includegraphics{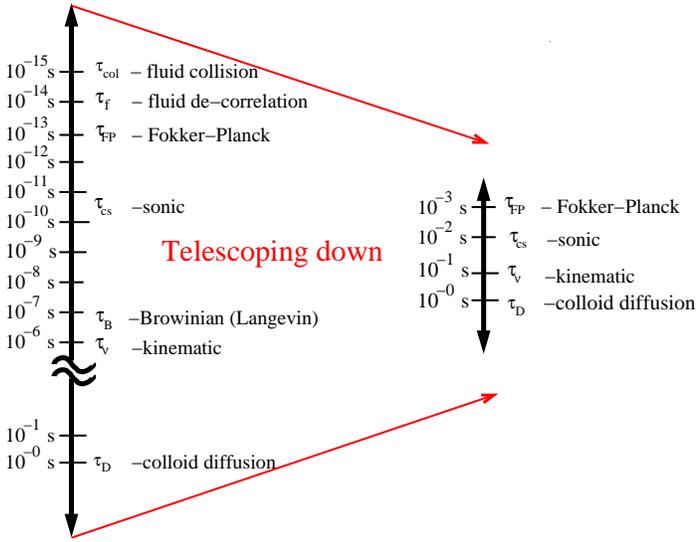}}
\caption{(Color online) Schematic depiction of our strategy for
coarse-graining across the hierarchy of time-scales for a colloid
(here the example taken is for a colloid of radius 1 $\mu
m$ in H$_2$0).  We first identify the relevant
time-scales, and then telescope them down to a hierarchy which 
is compacted to maximise simulation efficiency, but sufficiently separated
to correctly resolve the underlying physical behaviour.
\label{fig:Telescoping}
}
\end{figure}

 Finally, for the correct hydrodynamics to emerge we require that $
\tau_{FP},\tau_f \ll \tau_\nu \ll
\tau_D$.  But this separation no longer needs to be over many orders of
magnitude. As argued in ~\cite{Padd04}, one order of magnitude
separation between time-scales should be sufficient for most
applications.  We illustrate this approach in Fig.~\ref{fig:Telescoping}.

In the next few subsections we discuss in more quantitative detail how
our coarse-graining strategy telescopes down the hierarchy of
time-scales in order to maximize the efficiency of a simulation while
still retaining key physical features.

\subsubsection{SRD fluid time-scales}

In SRD the physical time-scale $\tau_{\mathrm{col}}$ is coarse-grained out, and
the effect of the collisions calculated in an average way every
time-step $\Delta t_c$.  The time-scale $\tau_f$ on which the 
velocity correlations decay can be quite easily
calculated from a random-collision approximation. Following
~\cite{Ripo05}: $\tau_f \approx
- \lambda t_0/\ln [1-\frac23(1 -
\cos \alpha )(1-1/\gamma)]$. For
our parameters, $\alpha=\frac12 \pi,
\gamma=5$,  we find $\tau_f \approx 0.76 \Delta t_c = 0.076 t_0$.
However, it should be kept in mind that for small $\lambda$ the
exponential decay with $t/\tau_f$ turns over to a slower algebraic decay
 at larger $t$~\cite{Ripo05}.  The amplitude of this ``tail'' is, however, quite small.

\subsubsection{Hydrodynamic time-scales for simulation}

For our choice of units $c_s = \sqrt{5/3}\,\, a_0/t_0$, so 
that the sonic time of Eq.~(\ref{eq:tcs}) reduces to
\begin{equation}
t_{cs} \approx 0.775 \frac{\sigma_{cf}}{a_0},
\end{equation}
which is independent of $\lambda$ or $\gamma$.

In the limit of small $\lambda$, the ratio of the kinematic time
$\tau_\nu$ to $\Delta t_c$ can be simplified to the following form:
\begin{equation}\label{eq:ratio-taunu}
\frac{\tau_\nu}{\Delta t_c} \approx 18 \frac{\sigma_{cf}^2}{a_0^2}
\end{equation}
 so that the condition $\tau_f,\tau_{FP} \ll \tau_\nu$ is very easy to
 fulfil.  Furthermore, under the same approximations, the ratio
 $\tau_\nu/t_{cs} \approx 23 \sigma_{cf}
\lambda$ so that for $\lambda$ too small the kinematic time becomes faster
than the sonic time.  For the simulation parameters used
in~\cite{Padd04} (and in Fig.~\ref{fig_vact_linearscale_enskog}), we
find $\tau_\nu/\tau_{cs} \approx 5$.

\subsubsection{Brownian  time-scales for simulation}

\begin{figure}[t]
  \scalebox{0.50}{\includegraphics{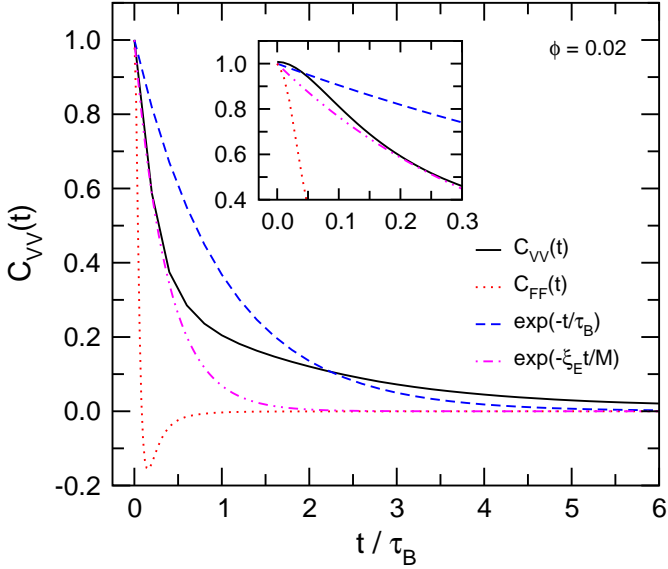}}
\caption{(Color online) Normalized VACF for a single colloid with $\sigma_{cf} = 2a_0$ and
$L=32 a_0$: measured (solid line),
Enskog short time prediction (dot-dashed line), and Brownian
approximation (dashed line).  The normalized force-force
auto-correlation $C_{FF}(t)$ (dotted line) decays on a much faster time scale.  The
inset shows a magnification of the short-time regime.  Note that the colloid
exhibits ballistic motion on a time-scale $\sim \tau_f$.
The time-scales $\tau_{FP} \approx 0.09$, $\tau_B \approx 2.5$ and $\tau_D \approx 200$ are 
clearly separated by at least an order of magnitude, as required.
\label{fig_vact_linearscale_enskog}
}
\end{figure}

We expect the Fokker Planck time $\tau_{FP}$ to scale as
$\Delta  t_c$, since this is roughly equivalent to the time $\tau_f$ over which the
fluid velocities will have randomized.  In Fig.~\ref{fig_vact_linearscale_enskog}
we plot the 
force-force correlation function
\begin{equation}\label{eq:force-force}
C_{FF}(t) = \frac{\left\langle F(t)F(0) \right\rangle}{\left\langle F^2 \right\rangle}.
\end{equation}
Its short time behavior is dominated by the random forces, and the
initial decay time gives a good estimate of $\tau_{FP}$. As can be seen
in the inset of Fig.~\ref{fig_vact_linearscale_enskog}, $\tau_{FP}
\approx 0.09 t_0$, which is indeed on the order of $\Delta t_c$ or
$\tau_f$.

If we make the reasonable  approximation that $\sigma_{cc} \approx 2 \sigma_{cf}$, then
the ratio of the Brownian time $\tau_B$ to the kinematic time $\tau_\nu$ 
simplifies to:
\begin{equation}\label{eq:ratio-tauB-taunu}
\frac{\tau_B}{\tau_\nu} \approx \frac{2\rho_c}{9\rho_f} 
\end{equation}
so that the ratio of $\tau_B$ to $\tau_\nu$ is  the same as for
a real physical system.  Since buoyancy requirements mean that
normally $\rho_c \approx \rho_f$, the time-scales are ordered as
$\tau_B < \tau_\nu$. Moreover, since the ratio $\tau_\nu/\Delta t_c \gg
1$, independently of $\gamma$ or even $\lambda$ (as long as $\lambda
\ll 1$), the condition that $\tau_{FP}
\ll \tau_B$ is also not hard to fulfil in an SRD simulation.

These time-scales are illustrated in
Fig.~\ref{fig_vact_linearscale_enskog} where we plot the normalized
time correlation function:
\begin{equation}
C_{VV}(t) = \frac{\left\langle V(t)V(0) \right\rangle}{\left\langle V^2 \right\rangle},
\end{equation}
with $V$ the Cartesian velocity coordinate of a single colloid.
After the initial non-Markovian quadratic decay, the subsequent short-time
exponential decay is not given by $\tau_B$ but instead by 
the {\em Enskog time}
\begin{equation}\label{eq:Enskog}
\tau_E = \frac{M_c}{\xi_E}
\end{equation}
 where $\xi_E$ is the Enskog friction which can be calculated from
 kinetic theory~\cite{Hyne77,Padd05}, see Eq.~(\ref{eq_xi_E}),
and generally $\tau_E < \tau_B$.
The physical origins of this behavior are described in more detail in 
Appendix~\ref{Langevin}.

The colloid diffusion coefficient is directly related to the friction by the 
Stokes-Einstein relation:
\begin{equation}\label{eq:Stokes-Einstein}
D_{\mathrm{col}} = \frac{k_B T}{\xi}.
\end{equation}
If we assume that $\lambda \ll 1$ and,  for simplicity, that $\xi \approx \xi_S$, i.e.\ we ignore the 
Enskog contribution, then the diffusion time scales as:
\begin{equation}\label{eq:diff-time}
\tau_D = \frac{\sigma_{cf}^2}{D_{\mathrm{col}}} \approx \frac{6 \pi \eta \sigma_{cf}^3}{k_B T} \approx \frac{\gamma}{\lambda} \left(\frac{\sigma_{cf}}{a_0}\right)^3 t_0 \approx  \frac{\gamma}{4 \lambda^2}\left(\frac{\sigma_{cf}}{a_0}\right) \tau_B,
\end{equation}
so that $\tau_B \ll \tau_D$ is not hard to fulfil.

It is also instructive to examine the ratio of  the diffusion time
$\tau_D$ to the kinematic time $\tau_\nu$:
\begin{equation}\label{eq:time-ratio2}
\frac{\tau_D}{\tau_\nu} = \frac{\nu}{D_{\mathrm{col}}}    \approx 0.06 \frac{\gamma}{\lambda^2}\left( \frac{\sigma_{cf}}{a_0 }\right).
\end{equation}
In general we advocate keeping $\lambda$ small to increase the Sc
number $\mbox{Sc}= \nu/D_f$, and since another obvious constraint is
$D_{\mathrm{col}} \ll D_f$, there is not too much difficulty achieving the desired separation of time-scales
 $\tau_\nu \ll \tau_D$.

As a concrete example of how these time-scales are separated in a
simulation, consider the parameters used in
Fig.~\ref{fig_vact_linearscale_enskog} for which  we find $\tau_f = 0.076
t_0$, $\tau_{FP} = 0.09 t_0$ $\tau_B = 2.5 t_0$, $\tau_\nu = 8 t_0$ and
$\tau_D = 200 t_0$.  But more generally, what the analysis of this section
shows is that obtaining the correct hierarchy of time-scales:
\begin{equation}
\tau_f , \tau_{FP} \ll \tau_E,\tau_B < \tau_\nu \ll \tau_D,t_S
\end{equation} 
is virtually guaranteed once the conditions on dimensionless numbers,
detailed in Section~\ref{Dimensionless}, are fulfilled.

\begin{figure}[t]
  \scalebox{0.50}{\includegraphics{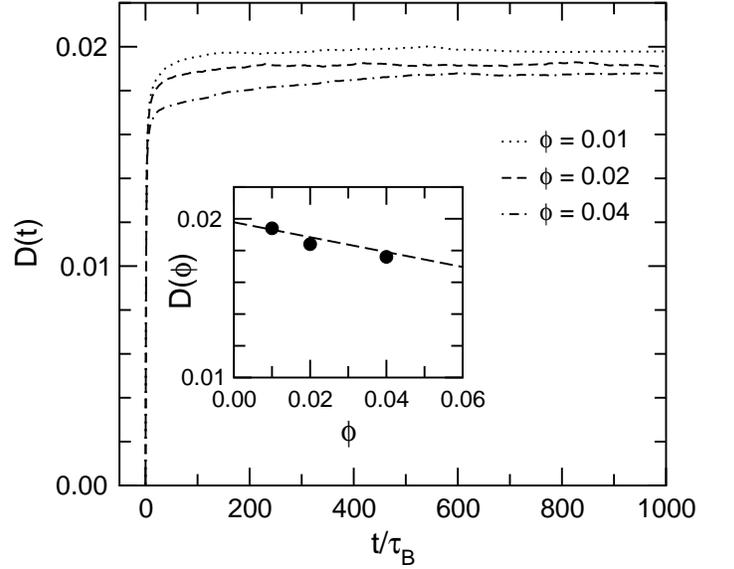}}
\caption{ 
Time dependent self-diffusion coefficient $D(t)$ for different volume
fractions $\phi$, achieved by varying the number of colloids in a box
of fixed volume. The inset shows the self-diffusion coefficient
($D=\lim t \rightarrow \infty D(t)$ as a function of volume fraction,
together with an analytical prediction $D = D_0 ( 1 - 1.1795
\phi)$~~\protect\cite{Cich88} valid in the low-density limit.
\label{fig_diffusion}
}
\end{figure}

\subsection{Measurements of diffusion}

To further test the hybrid MD-SRD coarse-graining scheme, we plot in 
Fig. \ref{fig_diffusion} the time-dependent diffusion
coefficient $D(t)$, defined as:
\begin{equation}\label{eq:Doft}
D(t) = \int_{0}^t \mathrm{d}t' \langle V(t')V(0) \rangle
\end{equation}
for a number of different volume fractions $\phi = \frac43 N/V \pi
\sigma_{cf}^3$.  Note that the convergence slows with
increasing volume fraction $\phi$. The infinite time integral gives the
diffusion coefficient, i.e.\ $D_{\mathrm{col}} = \lim_{t \rightarrow \infty} D(t)$.
In the limit of low densities the diffusion coefficient has been 
predicted to take the form  $D(\phi) = D_0 ( 1 - c
\phi)$ with
1.1795 for slip boundary spheres~\cite{Cich88}, and this provides a good fit
at the lower volume fractions.

\begin{figure}[t]
  \scalebox{0.50}{\includegraphics{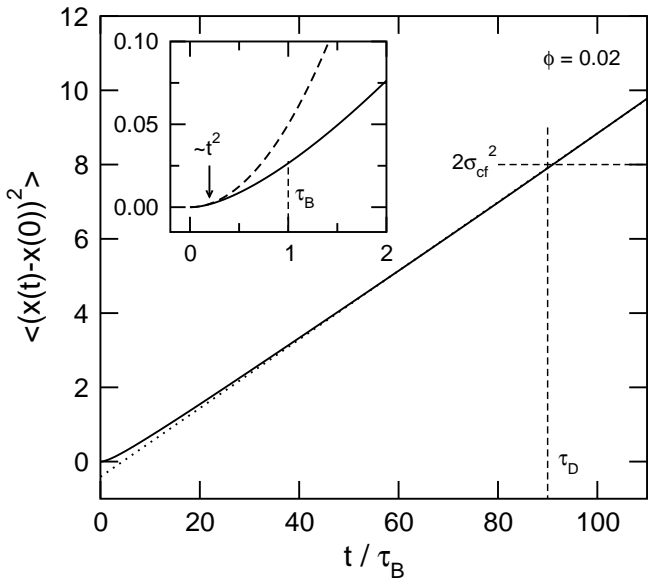}}
\caption{Mean-square-displacement of a $\sigma_{cf}=2 a_0$ colloid (solid line).
The dotted line is a fit to the long time limit.  The inset shows a
magnification of the short-time regime, highlighting the initial
ballistic regime (dashed line is $\langle V(0)\rangle^2 t^2$), which
rapidly turns over to the diffusive regime.
\label{fig_msd}
}
\end{figure}

In Figure \ref{fig_msd} we plot the mean-square displacement of a
Cartesian component of the colloidal particle position.  As
highlighted in the inset, at short times there is an initial ballistic
regime due to motion at an average mean square velocity $\langle
V(0)^2\rangle= k_B T/M$, resulting in a mean-square displacement
$\langle(X(t) -X(0))^2\rangle = (k_B T/M) t^2$. At times $t \gg
\tau_\nu$, the motion is clearly diffusive, as expected, with a linear
dependence on $t$ and a slope of $2 D_{\mathrm{col}}$, where
$D_{\mathrm{col}}$  is given by the infinite time limit of
Eq.~(\ref{eq:Doft}). On the scale of the plot the asymptotic regime is
reached well before the time $\tau_D$ over which the particle has
diffused, on average, over its radius.  In Appendix~\ref{Langevin} the 
behavior of $D(t)$ and the related mean-square displacement are discussed 
in more detail for time-scales $t \ll \tau_D$.

\section{Mapping between a physical and coarse-grained systems: No free lunch?}\label{mapping}

The ultimate goal of any coarse-graining procedure is to correctly
describe physical phenomena in the natural world.  As we have argued
in this paper, it is impossible to bridge, in a single simulation, all
the time and length-scales relevant to a colloid suspended in a
solvent.  Compromises must be made.  Nevertheless, a fruitful mapping
from a coarse-grained simulation to a physical system is possible, and
greatly facilitated by expressing the physical properties of interest
in dimensionless terms.  The best way to illustrate this is with some
examples.

\subsection{Example 1: Mapping to diffusive and Stokes time-scales}

For many dynamic phenomena in colloidal dispersions, the most
important time-scale is the diffusion time $\tau_D$.  To map a
coarse-grained simulation onto a real physical system one could
therefore equate the diffusion time and the colloid radius $a$ of the
simulation to that of the real physical system.  For example, take the
simulation parameters from table~\ref{tablecolloidSRD} in
Appendix~\ref{comparison} for $a = 2 a_0$, and compare these to the
properties of colloids of radius $a=1 \mu$m and $a=10$ nm from
table~\ref{tablecolloidwater}.  For the larger colloids, $\tau_D =
5$s, and equating this to the simulation $\tau_D$ means that $a_0= 0.5
\mu$m and $t_0 = 0.02$ s.  Performing the same exercise for the
smaller colloid, where $\tau_D = 5 \times 10^{-6}$ s, results in $a_0
= 5$ nm and $t_0 = 2 \times 10^{-8}$ s.  The first thing that becomes
obvious from this exercise is that ``physical'' time is set here not
by the coarse-grained simulation, but by the particular system that
one is comparing to.  In other words, many different physical systems
could be mapped to the same simulation.

If the system is also subjected to flow, then a second time-scale
emerges, the Stokes time $t_S = \mathrm{Pe} \,\,\tau_D$. As
long as the physical Pe number is achievable without compromising the
simulation quality, then this time-scale can simultaneously be set
equal to that of the physical system.  Behavior that depends
primarily on these two time-scales should then be correctly rendered by the
SRD hybrid MD scheme described in this paper.

\subsection{Example 2: Mapping to kinematic time-scales}

By fixing $\tau_D$ or $t_S$, as in example 1, other time-scales are
not correctly reproduced.  For the large and small colloid systems
described above, we would find $\tau_\nu = 0.17$~s and $1.7 \times
10^{-7}$~s respectively, which contrasts with the physical values of
$\tau_\nu = 10^{-6}$~s and $10^{-10}$~s shown in
Table~\ref{tablecolloidwater}.  In other words, fixing $\tau_D$
results in values of $\tau_\nu$ that are too large by orders of
magnitude. But of course these much larger values of $\tau_\nu$ are by
design, because the SRD simulation is optimized by compacting the very
large physical hierarchy of time-scales.  If we are interested in
processes dominated by $\tau_D$, having different $\tau_\nu$ doesn't
matter, so long as $\tau_\nu \ll \tau_D$.

On the other hand, if we want to describe processes that occur on
much smaller time-scales, for example long-time tails in colloidal
VACFs, then simulation times $\tau_\nu$ could be mapped onto the
physical $\tau_\nu$ instead. For the same $a=2 a_0$ simulation
parameters used above, the $a=1 \mu$m system now maps onto $a_0 = 0.5
\mu$m, and $t_0 = 1.3 \times 10^{-7}$~s with of course a strongly
underestimated diffusion time: $\tau_D = 3\times 10^{-5}$~s instead of
$5$~s.  Similarly for the $a=10$ nm system, the mapping is to $a_0 =
5$ nm, $t_0 =1.3 \times 10^{-11}$~s and $\tau_D = 3 \times 10^{-9}$~s
instead of $5 \times 10^{-6}$~s. For many processes on time-scales $t
< \tau_D$, this underestimate of $\tau_D$ doesn't really matter.  It
is only when time-scales are mixed that extra care must be employed in
interpreting the simulation results. And, if the processes can be
described in terms of different dimensionless times, then it should
normally still be possible to disentangle the simulation results and
correctly map onto a real physical system.

Another lesson from these examples of mapping to different time-scales
comes from comparing the kinematic viscosity of a physical system,
which say takes the value $\nu = 10^{6} \mu$m$^2$/s for water, to its
value in SRD simulations.  If for the $a=2a_0$ SRD system described
above we map time onto $\tau_D$, as in Example 1, then for the
$a=1\mu$m system $\nu =6 \mu$m$^2$/s, while for the $a=10$ nm system
$\nu = 6 \times 10^{-3}\mu$m$^2$/s.  As expected, both are much
smaller than the true value in water.  If we instead map the times to
$\tau_\nu$, then of course the kinematic viscosity takes the same
value as for the physical system (since we also set the length-scales
equal to the physical length-scales).  This apparent ambiguity in
defining a property like the kinematic viscosity is inherent in many
coarse-graining schemes.  We take the view that the precise value of
$\nu$ is not important, the only thing that matters is that it is
large enough to ensure a proper separation of times-scales and the
correct regime of hydrodynamic numbers.

\subsection{Example 3: Mapping mass and temperature}

In the examples above, we have only discussed setting lengths and
times.  This leaves either the mass $m_f$ or temperature (thermal
energy) $k_B T$ still to be set. Because for many dynamic properties
the absolute mass is effectively an irrelevant variable (although
relative masses may not be) there is some freedom to choose.  One
possibility would be to set $M_c$, the mass of the colloid, equal to
the physical mass.  For an $a=1 \mu$m neutrally buoyant colloid this
sets $m_f = 2.5 \times 10^{-14}$ g, (equal to about $8 \times 10^{8}$
water molecules) while for an $a=10$ nm colloid we find $m_f = 2.5
\times 10^{-20}$ g (equal to about $800$ water molecules). By
construction this procedure produces the correct physical $\rho_f$.
If we fix time with $\tau_\nu$, this will therefore also generate the
correct shear viscosity $\eta$. If instead we fix $\tau_D$, then
$\eta$ will be much smaller than the physical value. For either choice
of time-scales, other properties, such as the fluid self-diffusion
constant or the number density, will have different values from the
underlying physical system.

 Setting the mass in this way means that the unit of thermal energy $k_B
 T_0 = m_f a_0^2t_0^{-2}$ would also be very large, leading to the
 appearance of very low temperatures.  However, because temperature
 only determines behavior through the way it scales potentials, it is
 quite easy to simply re-normalize the effective energy scales and
 obtain the same behavior as for the physical system.  Alternatively
 one could set the temperature equal to that of a physical system, but
 that would mean that the masses would again differ significantly from
 the real physical system.  At any rate, for the same simulation (say
 of sedimentation at a given value of Pe) the values of mass or
 temperature depend on the physical system one compares to. This
 apparent ambiguity (or flexibility) simply reflects the fact that the
 dominant effects of temperature and mass come into play as relative
 ratios and not as absolute values.

\subsection{Example 4: Mapping to attractive potentials: problems with length-scales?}

So far we have only treated one length-scale in the problem, the
radius $a$.  Implicitly we have therefore assumed that the
colloid-colloid interaction does not have another intrinsic
length-scale of its own.  For hard-sphere colloids this is strictly
true, and for steep-repulsions such as the WCA form of
Eq.~(\ref{eq:col-col}) it is also a fairly good assumption that the
exact choice of the exponent $n$ in this equation does not
significantly affect the dynamical properties~\cite{Hans86}.
Similarly, there is some freedom to set the repulsive fluid-colloid
potential of Eq.~(\ref{eq:col-fluid}) in such a way as to optimize the
simulation parameters, in particular $\Delta t_{MD}$.  The physical
liquid-colloid interaction would obviously have a much more complex
form, but SRD coarse-grains out such details.  A similar argument can
be made for colloid and fluid interactions with hard walls, needed,
for example, to study problems with confinement, for which, {\it inter
alia}, SRD is particularly well suited.

Things become more complicated for colloids with an explicit
attractive interaction, such as DLVO or a depletion
potential~\cite{Russ89}.  These potentials introduce a new
length-scale, the range of the attraction. The ratio of this length to
the hard-core diameter helps determine the equilibrium properties of
the fluid~\cite{Loui01a,Liko01}.  In a physical system this ratio may
be quite small.  Keeping the ratio the same in the simulations then
leads to potentials that are very steep on the scale of $\sigma_{cc}$.
The MD time-step $\Delta t_{MD}$ may need to be very small in order to
properly integrate the MD equations of motion.  Such a small $\Delta
t_{MD}$ can make a simulation very inefficient, as was found in
~\cite{Hech05} who followed this strategy and were forced to use
$\Delta t_{MD}/\Delta t_c \approx 455$.  However, in that case the
DLVO potential dominates the behaviour, so that SRD was only included
to roughly resolve the long range hydrodynamics\cite{Hecht-comment}. In
general, we advocate adapting the potential to maximize simulation
efficiency, while preserving key physical properties such as the
topology of the phase diagram.  For example, the attractive energy
scale can be set by the dimensionless reduced second virial
coefficient, which gives an excellent approximation for how far one is
from the liquid-liquid critical point~\cite{Vlie00}.  When the
attractive interaction range is less than about $30 \%$ of the colloid
hard-core diameter, the colloidal ``liquid'' phase becomes metastable
with respect to the fluid-solid phase-transition.  At low colloid
packing fraction $\phi_c$ this leads to so-called ``energetic fluid''
behavior~\cite{Loui01a}.  To simulate a colloidal suspension in this
``energetic fluid'' regime, having a range of less than $30\%$ of $a$
should suffice.  In this way adding attractions does not have to make
the simulation much less efficient. Moreover the effects of these
constraints, placed by the colloid-colloid interaction on $\Delta
t_{MD}$, are to some degree mitigated by the fact that it is usually
the colloid-fluid interaction which sets this time-scale.

There are still a few final subtleties to mention.  First of all,
using small $\sigma_{cf}/a_0$ in the simulations may greatly increase
efficiency, but it also leads to a relatively larger contribution of
the Enskog friction to the total friction in Eq.~(\ref{eq_xi}).  For
physical colloids in a molecular solvent the exact magnitude of the
Enskog friction is still an open question, but undoubtedly for larger
colloids it is almost negligible.  Some of the Enskog effects can
simply be absorbed into an effective hydrodynamic radius $a$, but
others must be interpreted with some care.

Another regime where we might expect to see deviations beyond a simple
re-scaling of time-scales would be at very short times.  For example,
when colloids form a crystal, their oscillations can be de-composed
into lattice phonons.  All but the longest wave-length longitudinal
oscillations are overdamped due to the viscous drag and backflow
effects of the solvent~\cite{Hurd82}.  In SRD, however, some of the
very short wavelength oscillations might be preserved due to the fact
that the motion on those time-scales is still ballistic.  These may have 
an impact on the interpretation of  SRD simulations of microrheology.

In summary, these examples demonstrate that as long as the
coarse-grained simulation produces the correct hydrodynamic and
Brownian behavior then there is considerable freedom in assigning real
physical values to the parameters in the simulation.  Attractive
interactions may introduce a new length-scale, but a judicious choice
of a model potential should still reproduce the correct physical
behavior.  The same simulation may therefore be mapped onto multiple
different physical systems.  Conversely, for the same physical system,
different choices can be made for the time-scales that are mapped to
physical values, but by design, not all time-scales can be
simultaneously resolved.

\section{Conclusion}

Correctly rendering the principal Brownian and hydrodynamic properties
of colloids in solution is a challenging task for computer
simulation.  In this article we have explored how treating the solvent
with SRD, which coarse-grains the collisions between solvent particles
over both time {\em and} space, leads to an efficient solution of the
thermo-hydrodynamic equations of the solvent, in the external field
provided by the colloids.  Although it is impossible to simulate the
entire range of time-scales encountered in real colloidal suspensions,
which can span as much as 15 orders of magnitude, we argue that this
is also not necessary.  Instead, by recognizing firstly that
hydrodynamic effects are governed by a set of dimensionless numbers,
and secondly that the full hierarchy of time-scales can be telescoped
down to a much more manageable range while still being properly
physically separated, we demonstrate that the key Brownian and hydrodynamic
properties of a colloidal suspension can indeed be captured by our
SRD based coarse-graining scheme.

In particular, to simulate in the colloidal regime, the modeler
should ensure that the dimensionless hydrodynamic numbers such as the
Mach, Reynolds and Knudsen numbers are small enough ($\ll 1$), and the
Schmidt number is large enough ($\gg 1$).  While these numbers are
constrained by approximate limits, the Peclet number, which measures
the relative strength of the convective transport to the diffusive
transport of the colloids, can be either larger or smaller than 1,
depending on  physical properties such as the colloidal size and
the strength of the external field that one wants to study.  It turns
out that if the hydrodynamic numbers above are chosen correctly, it is
usually not so difficult to also satisfy the proper separation of the
time-scales.

We explored how to reach the desired regime of hydrodynamic numbers
and time-scales by tuning the simulation parameters. The two most
important parameters are the dimensionless mean-free path $\lambda$ which
measures what fraction of a cell size $a_0$ an SRD particle travels
between collisions performed at every time-step $\Delta t_c$, and the ratio of
the colloid radius $\sigma_{cf}$ to the SRD cell size $a_0$.  The
general picture that emerges is that the collision time $\Delta t_c$
must be chosen so that $\lambda$ is small since this helps keep the
Schmidt number high, and both the Knudsen and Reynolds numbers low.
Of course for computational efficiency, the collision time should not
be too small either.  Similarly although a larger colloid radius means
that the hydrodynamic fields are more accurately rendered, this should
be tempered by the fact that the simulation efficiency drops off
rapidly with increasing colloid radius.  We find that choosing
$\sigma_{cf}/a_0
\approx 2$, which may seem small, already leads to accurate
hydrodynamic properties.

A number of subtleties occur at the fluid-colloid interface, which may
induce spurious depletion interactions between the colloids, but these
can be avoided by careful choice of potential parameters for slip
boundary conditions.  Stick boundary conditions can also be
implemented, and we advocate using stochastic bounce-back rules to
achieve this coarse-graining over the fluid-colloid interactions.

The simplicity of the SRD solvent facilitates the calculation of the
the short-time behavior of the VACF from kinetic theory.  We argue
that for times shorter than the sonic time $t_{cs}=a/c_s$, where
hydrodynamic collective modes will not yet have fully developed, the
decay of the VACF is typically much more rapid than the prediction of
the simple Langevin equation.  The ensuing Enskog contribution to the
total friction should be added in parallel to the microscopic
hydrodynamic friction, and from the sum an analytic expression for the
effective hydrodynamic radius $a_{\mathrm{eff}}$ can be derived which
is consistent with independent measurements of the  long-ranged hydrodynamic fields.

Although we have shown how to correctly render the principal Brownian
and hydrodynamic behavior of colloids in solution there is, as always,
{\em no such thing as a free lunch}.  The price to be paid is an inevitable
consequence of compressing together the hierarchy of time and
length-scales: not all the simulation parameters can be simultaneously
mapped onto a physical system of interest.  However the cost of the
``lunch'' can be haggled down by recognizing that only a few physical
parameters are usually important.  For example, one could map the
simulation onto real physical time, say through $\tau_D$ and $t_S$, or
alternatively through $\tau_\nu$; there is some freedom (or ambiguity)
in the choice of which time scale to use.  The correspondence with
physical reality becomes more complex when different physical
time-scales mix, but in practice they can  often be disentangled when
both the coarse-grained simulation and the physical system are
expressed in terms of dimensionless numbers and ratios.  In other
words, there is no substitute for careful physical insight which is,
as always, priceless.

A number of lessons can be drawn from this exercise that may be
relevant for other coarse-graining schemes.  SRD is closely related to
Lattice-Boltzmann and to the Lowe-Anderson thermostat, and many of the
conclusions above should transfer in an obvious way.  For dissipative
particle dynamics, we argue that the physical interpretation is
facilitated by recognizing that, just as for SRD, the DPD particles
should not be viewed as ``clumps of fluid'' but rather as a convenient
computational tool to solve the underlying thermo-hydrodynamic
equations.  All these methods must correctly resolve the time-scale
hierarchy, and satisfy the relevant hydrodynamic numbers, in order to
reproduce the right underlying physics.

 Our measurements of the VACF and the discussion in Appendix B show
explicitly that Brownian Dynamics simulations, which are based on the
simple Langevin equation, do not correctly capture either the long or
the short-time decay of colloidal VACFs.

A major advantage of SRD is the relative ease with which solutes and
external boundary conditions (walls) are introduced.  Boundaries may
be hard or soft, with either stick or slip conditions. This method is
therefore particularly promising for simulations in the fields of bio-
and nanofluidics, where small meso particles flow in constrained
geometries, possibly confined by soft fluctuating walls.  We are
currently exploring these possibilities.

\acknowledgments
JTP thanks the EPSRC and IMPACT Faraday, and AAL thanks the Royal
Society (London) for financial support.  We thank W. Briels,
H. L\"{o}wen, J. Lopez Lopez, J. San\'e, A. Mayhew Seers,
I. Pagonabarraga, A. Wilber, and A. Wysocki for very helpful
conversations, and W. den Otter for a careful reading of the
manuscript.

\appendix

\section{Thermostating of SRD under flow}

When an external field is applied to the fluid (or to objects embedded
in the fluid), energy is pumped into the system and, as a consequence,
the average temperature will rise.  To prevent this from happening,
the system must be coupled to a thermostat.  In order not to influence
the average flow, thermostating requires a local and Galilean
invariant definition of temperature.

We achieve this by relating the instantaneous local temperature
in a cell to the mean square deviation of the fluid particle velocities from the center
of mass velocity of that cell. To minimize interference of the thermostat with
the dynamics, we choose to measure the overall temperature as:
\begin{eqnarray}
k_BT_{\mathrm{meas}} & = & \frac{m_f}{N_{\mathrm{free}}}
 \sum_{\mathrm{cell}\,c} \sum_{i \in c}
(\mathbf{v}_i - \mathbf{v}_{\mathrm{cm},c})^2 \\
N_{\mathrm{free}} & = & \sum_{\mathrm{cell}\,c} \left\{
\begin{array}{ll}
	3 (N_{\mathrm{cell}\,c}-1) & (N_{\mathrm{cell}\,c} > 1) \\
	0 & (N_{\mathrm{cell}\,c} \leq 1)
\end{array}
\right.
\label{eq_Nfree}
\end{eqnarray}
In Eq.~(\ref{eq_Nfree}), $N_{\mathrm{cell}\,c}$ is the instantaneous
number of fluid particles within cell $c$.  Three degrees
of freedom must be subtracted for fixing the center of mass velocity of each cell
(note that this implies that the local temperature is not defined in
cells containing 0 or 1 particle). During the simulations, the temperature
$T_{\mathrm{meas}}$ is measured every $\Delta t_c$ (this can be
done very efficiently because the relative velocities are readily available
in the collision step routine). The thermostat then acts by rescaling all relative
velocities $\mathbf{v}_i - \mathbf{v}_{\mathrm{cm},c}$ by a factor
$\sqrt{T/T_{\mathrm{meas}}}$. This strict enforcement of the overall temperature
may be relaxed by allowing appropriate fluctuations, but in view of the very
large number of fluid particles (typically $10^6$) this will not have a measurable
effect on the dynamics of the fluid particles.

\section{The Langevin equation and memory effects}\label{Langevin}

Within the Brownian approximation each Cartesian component of the
colloid velocity $V$ is described by a simple Langevin equation of the
form:
\begin{eqnarray} \label{eq:Langevin}
M_c \frac{\mathrm{d}V}{\mathrm{d}t} & = & - \xi V + F^R(t) \\
\left\langle F^R(t) F^R(t') \right\rangle & = & 2 k_BT \xi \delta(t-t')
\label{eq:memory}
\end{eqnarray}
without any memory effect in the friction coefficient $\xi$, or in the
random force $F^R(t)$.  This Langevin equation fundamentally arises from the
assumption that the force-kicks are Markovian: each step is
independent of any previous behavior.  Of course on a very short
time-scale this is not true, but since Brownian motion is
self-similar, the Markovian assumptions underlying
Eq.~(\ref{eq:Langevin}) might be expected to be accurate on longer
time-scales.  Inspection of Fig.~\ref{fig_vact_linearscale_enskog}
shows that the force-force correlation function~(\ref{eq:force-force}) does
indeed decay on a time-scale $\tau_{FP}$ which is much faster than
other Brownian time-scales in the system.

  From the Langevin Eq.~(\ref{eq:Langevin}) it follows that the
  velocity auto-correlation function (VACF) takes the form
\begin{equation}\label{eq:cvv}
\left\langle V(t)V(0) \right\rangle^L = \frac{k_BT}{M_c} \exp(-t/\tau_B)
\end{equation}
where $\tau_B = M_c/\xi$.
  The VACF is directly related
to the diffusion constant through the Green-Kubo relation:
\begin{equation}\label{eq:D}
D_{\mathrm{col}} = \lim_{t \rightarrow \infty} \int \mathrm{d}t \left\langle V(t)V(0)\right\rangle = \frac{k_B T}{\xi}
\end{equation}
and Eq.~(\ref{eq:cvv}) has the attractive property that its integral
gives the correct Stokes-Einstein form for $D_{\mathrm{col}}$, and
that its $t =0$ limit gives the expected value from equipartition.  It
is therefore a popular pedagogical device for introducing Brownian
motion.  Notwithstanding its seductive simplicity, we will argue that
this Langevin approach strongly underestimates  the initial  decay rate 
of the VACF, and badly overestimates its long-time decay rate.

\subsection{VACF at long times}

At longer times, as first shown in MD simulations by Alder and
Wainwright~\cite{Alde70}, the VACF shows an algebraic decay that is
much slower than the exponential decay of Eq.~(\ref{eq:cvv}).  The
difference is due to the breakdown of the Markovian assumption for
times $t \gg \tau_{FP}$.  Memory effects arise because the momentum
that a colloidal particle transfers to the fluid is conserved, with
dynamics described by the Navier Stokes equations, resulting in
long-ranged flow patterns that affect a colloid on much longer
time-scales than Eq.(\ref{eq:memory}) suggests.  Calculations taking
into account the full time-dependent hydrodynamics were performed by
Hauge~\cite{Haug73} and Hinch~\cite{Hinc75} (and others apparently
much earlier~\cite{Lisy04}).
For $\rho_c=\rho_f$, i.e.\ a neutrally buoyant particle, these
expressions can be written as
\begin{equation}\label{eq:Hinch}
\langle V(t) V(0)\rangle^{H} 
 = \frac{2}{3} \frac{k_B T}{M_c} \left( \frac{1}{3 \pi} \int_0^{\infty} \mathrm{d}x \frac{ x^{\frac12} \exp\left[-x(t/\tau_\nu)\right]}{1 + \frac{x}{3} + \frac{x^2}{9}} \right)
\end{equation}
where we have chosen to use an integral form as in ~\cite{Paul81}.
Under the assumption that the sound effects have dissipated away, the
only hydrodynamic time-scale  is the kinematic time $\tau_\nu$
defined in Eq.~(\ref{eq:taunu}).  The VACF must therefore scale  with
$t/\tau_\nu$, independently of colloid size.  Eq.~(\ref{eq:Hinch}) is
 written in a way to emphasize this scaling.

  By expanding
the denominator it is not hard to see that at longer times the
well-known algebraic tail results:
\begin{equation}\label{eq:tail}
\lim_{t \to \infty} \langle V(t) V(0)\rangle^{H} \approx \frac{k_B T}{M_c} \frac{ 1}{9 \sqrt{\pi}}\left(\frac{\tau_\nu}{t}\right)^{3/2} = \frac{k_BT}{12 m \gamma \left( \pi \nu t \right)^{3/2}}
\end{equation}
with the same pre-factor that was found from mode-coupling
theory~\cite{Erns70}.

  The integral of Eq.~(\ref{eq:Hinch}) is
\begin{equation}\label{eq:DH}
\int_0^{\infty} \mathrm{d}t \langle V(t) V(0)\rangle^{H} = \frac{k_B T}{6 \pi \eta a},
\end{equation}
which means that all the hydrodynamic contributions to the friction
$\xi$ come from $\langle V(t) V(0) \rangle^{H}$. The integral
converges slowly, as $(t/\tau_\nu)^\frac12$, because of the long-time
tail. The VACF can easily be related to the mean-square displacement,
and deviations from purely diffusive motion are still observable for
$t \gg \tau_\nu$, and have been measured in
experiments~\cite{Paul81,Weit89,Luki05}.

 Although the last equality in Eq.~(\ref{eq:tail}) shows that the VACF
 tail amplitude is independent of colloid size, it should be kept in
 mind that the algebraic decay does not set in until $t
\approx \tau_\nu = a^2/\nu$. Thus it is the relative (not absolute) effect of the tail that is the same for all colloids.
For example, at $t=\tau_\nu$, the VACF will have decayed to $4\%$ of its
initial value of $k_B T/M_c$, independently of colloid mass $M_c$.  Even
though this tail amplitude may seem small, about half the weight of
the integral~(\ref{eq:DH}) that determines the friction is from $t
\gtrsim \tau_\nu$.  The dominance of the non-Markovian hydrodynamic
behavior in determining the friction for colloids does not appear to
hold for microscopic fluids, where, for example for LJ liquids near
the triple point, it is thought that the long-time tail only adds a
few $\%$ to the friction coefficient~\cite{Leve74}.  However, this is
not due to the simple Langevin equation~(\ref{eq:Langevin}) being more
accurate, but rather to the modified short-time behavior of the VACF,
caused by fluid correlations (a different non-Markovian effect).

\begin{figure}[t]
  \scalebox{0.50}{\includegraphics{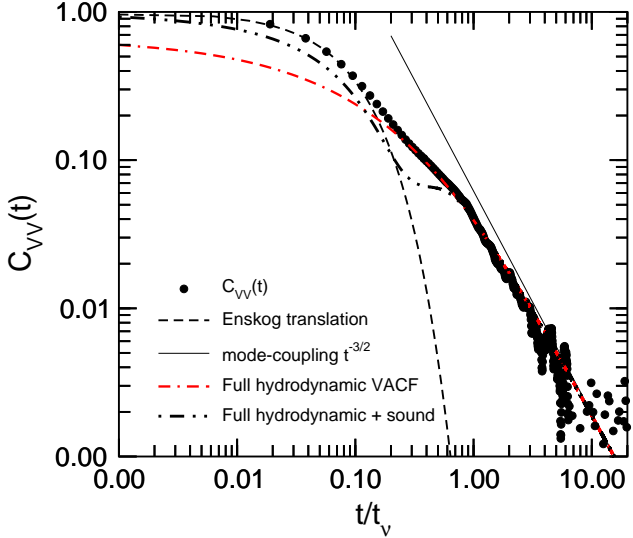}}
\caption{(Color online) Normalized VACF for $a=2 a_0$ and stick boundary conditions (from~\protect\cite{Padd05}),  compared to the short-time Enskog result~(\protect\ref{eq:cvvEnskog}),
and the full hydrodynamic VACF~(\protect\ref{eq:Hinch}) by itself, and 
with the  VACF~(\protect\ref{eq:cvvsound}) from sound  added.
We also show the asymptotic long-time hydrodynamic tail from mode-coupling theory
\label{fig:autocorrelations-A=2-logplot}
}
\end{figure}

\subsection{VACF at short times}

At short times the hydrodynamic contribution to the
VACF~(\ref{eq:Hinch}) reduces to
\begin{equation}
\lim_{t \to 0} \langle V(t) V(0)\rangle^{H} = \frac{2}{3} \frac{k_B T}{M_c}
\end{equation}
because, within a continuum description, $\frac13$ of the energy is
dissipated as a sound wave at velocity $c_s$,
\begin{eqnarray}\label{eq:cvvsound}
\langle V(t) V(0)\rangle^{cs} &= &
\frac{1}{3} \frac{k_B T}{M_c} \exp\left[-(3/2)(t/t_{cs})\right] \cdot \nonumber \\
& & \left[ \cos\left( \frac{\sqrt{3}}{2} \frac{t}{t_{cs}}\right) -
\sqrt{3} \sin\left( \frac{\sqrt{3}}{2} \frac{t}{t_{cs}}\right) \right]
\nonumber \\
\end{eqnarray}
where the sonic time $t_{cs}=a/c_s$ is defined in Table~\ref{table:time}.
The sound wave doesn't contribute to the friction or diffusion since
 \begin{equation}
\int_0^{\infty} \mathrm{d}t \langle V(t) V(0)\rangle^{cs} = 0.
\end{equation}
For a more extended discussion of the role of sound on hydrodynamics see e.g.\ ref.~\cite{Bakk02}.

Because $t_{cs} \ll \tau_\nu$, the sound-wave contribution to the VACF
decays much faster than the hydrodynamic contribution
 $\left\langle V(t)V(0) \right\rangle^H$ or
even than the Langevin approximation $\left\langle V(t)V(0)
\right\rangle^L$ (recall that $\tau_\nu = \frac92 \tau_B$ for neutrally buoyant
colloids.)  Therefore, in the continuum picture, the short-time decay
of the VACF is much faster than that suggested by the simple Langevin
equation.  And even after the sound wave has decayed, $\left\langle V(t)V(0)
\right\rangle^H$ decays much more rapidly (faster than simple exponential) than Eq.~(\ref{eq:cvv}) for times $t \lesssim \tau_B$.  SRD simulations confirm this more rapid  short time decay.  For
example, in Fig~\ref{fig:autocorrelations-A=2-logplot} we show a
direct comparison of the measured VACF to the continuum approach,
where $\left\langle V(t)V(0) \right\rangle = \left\langle V(t)V(0)
\right\rangle^H + \left\langle V(t)V(0) \right\rangle^{cs}$.  For
short times the decay predicted from sound agrees reasonable well, and
for long times the full hydrodynamic theory quantitatively fits the
simulation data.

A closer look at the short-time behavior in our SRD simulations, both 
for slip boundaries as in Fig.~\ref{fig_vact_linearscale_enskog} 
and for stick-boundaries as in Fig~\ref{fig:autocorrelations-A=2-logplot} 
or in~\cite{Padd05}, shows that a better fit 
to the short-time exponential decay of the VACF is given by
\begin{equation}\label{eq:cvvEnskog}
\langle V(t) V(0)\rangle^{E} = \frac{k_B T}{M_c} \exp\left(-t/\tau_E\right),
\end{equation} 
where the Enskog time~(\ref{eq:Enskog}) scales
as:
\begin{equation}\label{eq:Enskog-scaling}
\tau_E = \frac{M_c}{\xi_E} \approx \frac {0.5\,a}{\sqrt{k_B T/m}} \approx 0.645 t_{cs}
\end{equation}
for heavy (stick boundary) colloids in an SRD fluid.  Note that the Enskog decay time
is very close to the decay time $\frac23 t_{cs}$ of the sound
wave~(\ref{eq:cvvsound}).  That the SRD Enskog time should scale as
$t_{cs}$ is perhaps not too surprising, because that is the natural
time-scale of the SRD fluid.  However, the fact that the pre-factors
in the exponential of Eqs~(\ref{eq:cvvsound}) and~(\ref{eq:cvvEnskog})
are virtually the same may be accidental. Although in a real liquid
the increased compressibility factor means $\tau_E$ and $\tau_{cs}$
are lower than their values for an ideal gas, the two times  change
at slightly different rates so that we don't expect the same 
equivalence in physical systems.

In contrast to the sound wave, the integral over the Enskog
VACF leads to a finite value:
\begin{equation}\label{eq:Enskog-integral}
\int_0^{\infty} \mathrm{d}t \langle V(t) V(0)\rangle^{E} = \frac{k_B T}{\xi_E}
\end{equation}
which quantitatively explains the friction we measured in our
simulations, as discussed in section~\ref{finite-size}.  In contrast
to the VACF from sound, the Enskog result cannot simply be added to
the VACF from Eq.~(\ref{eq:Hinch}) because this would violate the
$t=0$ limit of the VACF, set by equipartition.  However, because the
Enskog contribution occurs on times $t \lesssim t_{cs}$, where sound
and other collective modes of (compressible) hydrodynamics are not yet
fully developed, it may be that Eqs.~(\ref{eq:Hinch})
and~(\ref{eq:cvvsound}), derived from continuum theory, should be
modified on these short time-scales.  Since the dominant contribution
to the integral in Eq.~(\ref{eq:DH}) is for longer times, the exact
form of the hydrodynamic contribution~(\ref{eq:Hinch}) on these short
times ($t
\lesssim t_{cs}$) does not have much influence on the hydrodynamic
part of the friction.  Thus the approximation of the short-time VACF
by the Enskog result of Eq.~(\ref{eq:cvvEnskog}), and the longer-time
($t \gg \tau_E$, but still $t < \tau_\nu)$) VACF by Eq.~(\ref{eq:Hinch})
may be a fairly reasonable empirical approach.  This approximation,
together with the Green-Kubo relation~(\ref{eq:D}), then justifies,
a-posteori, the parallel addition of frictions in Eq.~(\ref{eq_xi}).

Just as was found for the long-time behavior, the short-time behavior
of the VACF displays important deviations from the simple Langevin
picture. We find that the VACF decays much faster, on
the order of the sonic time $\tau_{cs}$ rather than the Brownian time
$\tau_B$.  From the tables
\ref{tablecolloidwater} and
\ref{tablecolloidSRD} one can see how large these differences can be.
For example, for colloids with $a= 1 \mu m$, the short-time Enskog or
sonic decay times are of order $10^{-9}$ s, while $\tau_B$ is of order
$10^{-7}$ s, so that the decay of the VACF will be dominated by
$<V(t)V(0)>^H$, with its non-exponential behavior, over almost the entire
time regime.

\begin{figure}[t]
  \scalebox{0.50}{\includegraphics{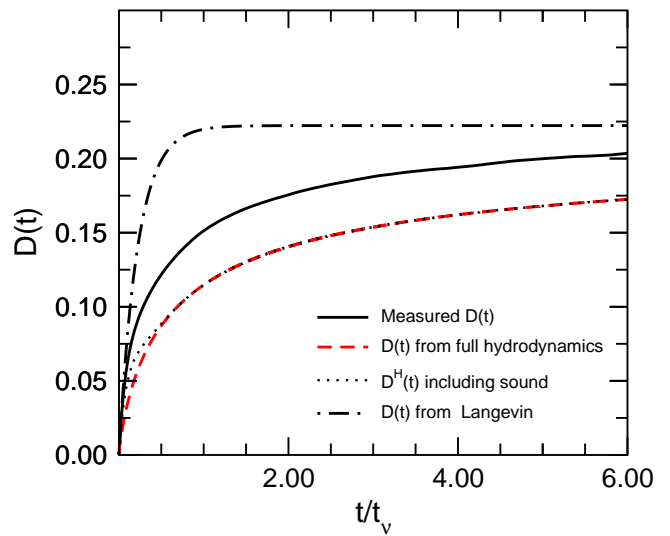}}
\caption{(Color online) Integral $D(t)$ of the normalized VACF from 
Fig.~\protect\ref{fig:autocorrelations-A=2-logplot} (solid line), 
compared to the integral of Eq.~(\protect\ref{eq:Hinch}) (dashed
line), and including the contribution from sound as in
Eq.~(\protect\ref{eq:DofH}) (dotted line). The simple Langevin
form from the integral of Eq.~(\protect\ref{eq:cvv}) (dash-dotted line), is calculated 
with $\xi = 6 \pi \eta
\sigma_{cf}$ in order to have the same asymptotic value as $D^H(t)$.  Because in the plot we integrate the normalized VACF, 
the $t \rightarrow \infty$ limit of $D(t)$ for the two theoretical
approaches is $\frac29t_0$ in our units, independent of colloid size.
The measured $D(t)$ has a larger asymptotic value due to the initial
effect of the Enskog friction.
\label{fig:Diffusion-A=2-linplot}
}
\end{figure}

An example of the effect of the Enskog contribution on the friction or
the mean-square displacement is given in
Fig.~\ref{fig:Diffusion-A=2-linplot}. The integral of the VACF, $D(t)$
from Eq.~(\ref{eq:Doft}), which is directly related to the mean-square
displacement~\cite{Hans86}, is extracted from the simulations described in
Fig~\ref{fig:autocorrelations-A=2-logplot}, and compared to
\begin{equation}\label{eq:DofH}
D^H(t) =  
\int_0^t \mathrm{d}t'\left(  \left\langle V(t')V(0)
\right\rangle^H + \left\langle V(t')V(0) \right\rangle^{cs}\right) 
\end{equation}
calculated using Eqs.~(\ref{eq:Hinch})~and~(\ref{eq:cvvsound}).  Both
show exactly the same $(t/\tau_\nu)^\frac12$ scaling for $t \gtrsim \tau_\nu$, but
the simulated $D(t)$ is larger because  the initial Enskog
contribution means that $D(t) \approx D^H(t) + D_E$,
for $ t \gg \tau_E$, with
$D_E = k_BT /\xi_E$ defined in Eq.~(\ref{eq:Enskog-integral}).

 The slow algebraic convergence to the final diffusion coefficient, or
 related friction, also helps explain the scaling of the finite size
 effects of Eq.~(\ref{eq:f}).  A finite box of length $L$ means that
 the measured $D(t)$ is cut off roughly at $t_L = (L^2/a^2) \tau_\nu$,
 and because $\int dt D(t) \sim (\tau_\nu/t)^{\frac12}$ for $t \gtrsim \tau_\nu$ the effect on the
 diffusion constant can be written as $D \approx D_0(1 - c (L/a))$
 which explains  the scaling form of Eq.~(\ref{eq:f}).
 Fig.~\ref{fig:Diffusion-A=2-linplot} also demonstrates how much more 
rapidly the $D(t)$
 from the Langevin Eq.~(\ref{eq:cvv}) converges to its final result $D = k_B T/\xi$ 
  when compared to the simulations and hydrodynamic theories.

In summary, even though the simple Langevin
equation~(\ref{eq:Langevin}), without memory, may have pedagogical
merit, it does a poor job of describing the VACF for colloids in
solution.

\section{Comparison of SRD to physical parameters}\label{comparison}

At the end of the day one would like to compare a coarse-grained
simulation to a real physical colloidal dispersion.  To help
facilitate such comparisons, we have summarized a number of physical
parameters and dimensionless numbers for colloids in water, and also
colloids in an SRD fluid. 

 Table~\ref{tableSRD} compares the properties of a pure SRD fluid, in
 the limit of small mean-free path $\lambda$, to the properties of
 water.

 Table~\ref{tablecolloidwater} summarizes the main scaling and values
 for a number of different properties of two different size neutrally buoyant
 colloids in water.  To approximate the Enskog friction we used a
 simplified form valid for stick boundary conditions in an ideal
 fluid, as in~\cite{Padd05}, which ignores the compressibility and
 other expected complexities of real water. These predictions should therefore 
be taken with a grain of salt.  For simplicity, we have
 ignored any potential Enskog effects on the diffusion constant or
 time-scales.

Table~\ref{tablecolloidSRD} shows a very similar comparison for two
different size colloids in an SRD fluid.  For simplicity we assume
that there are no finite-size effects, which explains small
differences when compared to similar parameters used in the text of
the main article.  However, in contrast to
Table~\ref{tablecolloidwater}, we do include the Enskog contribution
when calculating the total friction and related properties.

\begin{table*}
\begin{center}

\caption{Summary of the physical properties of 
a 3D SRD fluid in the regime of small reduced mean-free path
$\lambda$.  In the first column we show how these parameters scale for
SRD. To highlight the scaling with the reduced mean-free path$
\lambda$ and the number of particles per cell $\gamma$, we only show
the collisional contribution to the viscosity in
Eqs.~(\protect\ref{eq:nukin})~-~(\protect\ref{eq:etacol}), which
dominates for small $\lambda$, and also ignore the factor
$f_{\mathrm{col}}^\nu(\gamma,\alpha)$.  In the second column we list
the parameter values for our choice of simulation parameters in this
paper and in refs~\protect\cite{Padd04,Padd05}, and in the last column
we compare to the parameter values for H$_2$0 at standard temperature
and pressure.  For the SRD the units are in terms of $m_f$, $a_0$ and
$k_BT$, so that the time unit is $t_0 = a_0\sqrt{m_f/k_B T}$. 
\label{tableSRD}
 }
\begin{ruledtabular}
\begin{tabular}{llccc}
\,& \, &  scaling in  SRD  &  value for $\lambda = 0.1$, $\gamma=5$, $\alpha = \frac12 \pi$ &  value for H$_2$0 \\  \hline \noalign{\medskip} 

$\rho_f$ & mass density & $\displaystyle \gamma \,\,\frac{m_f}{a_0^3}$
& $\displaystyle 5 \,\,\frac{m_f}{a_0^3}$
 & $\displaystyle 1\times 10^{-12} \,\, \frac{\mathrm{g}}{\mu \mathrm{m}^3}$ \\
\noalign{\medskip}

$n_f$ & number density & $\displaystyle\gamma \,\, a_0^{-3} $ &
$\displaystyle 5 \,\, a_0^{-3} $ &
$\displaystyle 3.35 \times 10^{10} \,\, \mu \mathrm{m}^{-3} $\\
\noalign{\medskip}

$c_s$ & speed of sound & $\displaystyle \sqrt{\frac{5k_B T}{3m_f}}
  $ &
$\displaystyle 1.29\,\,\frac{a_0}{t_0} $ &
 $\displaystyle 1.48 \times 10^{9}
\,\, \frac{\mu \mathrm{m}}{\mathrm{s}} $ \\ \noalign{\medskip}

$\lambda_\mathrm{free}$ & mean free path & $\lambda_\mathrm{free} = \lambda a_0  $ &
$ 0.1 a_0
$ &
 $ \approx 3$ \AA  \\ \noalign{\medskip}

 $\eta_f$ & shear viscosity & $\displaystyle\approx \frac{\gamma}{18
 \lambda} \,\, \frac{m_f}{a_0 t_0}$ &
$\displaystyle\approx 2.50 \,\, \frac{m_f}{a_0 t_0}$ &
 $\displaystyle 1 \times 10^{-6}
 \, \frac{ \mathrm{g}}{\mu \mathrm{m} \,\mathrm{s}}$ \\ \noalign{\medskip}

$\nu$ & kinematic viscosity & $\displaystyle\approx \frac{1}{18
\lambda}\,\, \frac{a_0^2}{t_0}$ &
 $\displaystyle\approx 0.50 \,\, \frac{a_0^2}{t_0}$ &
 $\displaystyle 1
\times 10^{6} \,\, \frac{\mu \mathrm{m}^2}{\mathrm{s}}$
\\ \noalign{\medskip}

$D_f$ & self diffusion constant & $\displaystyle \approx \lambda \,\,
\frac{a_0^2}{t_0}$ &
$\displaystyle \approx 0.1 \,\,
\frac{a_0^2}{t_0}$ &
 $\displaystyle 2.35
\times 10^{3} \,\, \frac{\mu \mathrm{m}^2}{\mathrm{s}} $ \\ \noalign{\medskip}

$\mathrm{Sc} \,\,$ & Schmidt number & $ \displaystyle \frac{\nu}{D_f} \approx
\frac{1}{18 \lambda^2} $ & 
 $ \displaystyle \frac{\nu}{D_f} \approx
5 $ & 
$ \displaystyle \frac{\nu}{D_f} \approx 425 $

\\ \noalign{\medskip}

\end{tabular}
\end{ruledtabular}
\end{center}
\end{table*}

\begin{table}
\begin{center}
\caption{Physical parameters, dimensionless hydrodynamic numbers and
 time-scales for spherical colloids of radius $a=0.01 \mu \mathrm{m}$ ($10 nm$) and $a= 1
 \mu \mathrm{m}$ in H$_2$0 at standard temperature and pressure, moving at a
 velocity $v_S = 10\mu \mathrm{m}/\mathrm{s}$.  The mass density $\rho_c$ of the colloid
 is taken to be the same as that of water (the colloid is neutrally
 buoyant), and the hydrodynamic radius $a$ is taken to be the same as
 the physical radius.  Where, for notational clarity, the units are
 not explicitly shown, then length is measured in $\mu \mathrm{m}$, time in
 s, and mass in g.
\label{tablecolloidwater}
 }
  
\begin{ruledtabular}
\begin{tabular}{lll}
Physical parameters &  $a=10 n\mathrm{m}$ & $a=1 \mu \mathrm{m}$ \\  \noalign{\medskip} \hline \noalign{\medskip}

$\displaystyle M  \approx \frac{4}{3} \pi a^3 \times 10^{-12} \frac{\mathrm{g}}{\mu \mathrm{m}^3} \,  $& $\displaystyle
 4.19 \times 10^{-18} \mathrm{g} $  & $\displaystyle 4.19 \times 10^{-12} \mathrm{g}$  \\\noalign{\medskip}

$\displaystyle \frac{\xi_E}{\xi_S}  \approx  1.6 \times 10^{2} a \,  $& $\displaystyle
 \approx  1.6 $  & $\displaystyle  \approx 1.6 \times 10^{2}$  \\\noalign{\medskip}

$\displaystyle \xi_S = 6 \pi \eta a \approx 1.9 \times 10^{-5} a $&
$\displaystyle \approx 1.9 \times 10^{-7} \,\, \frac{\mathrm{g}}{\mathrm{s}} $ &
$\displaystyle \approx 1.9 \times 10^{-5} \,\, \frac{\mathrm{g}}{\mathrm{s}} $
\\\noalign{\medskip}

$\displaystyle D_{\mathrm{col}} \approx  \frac{k_B T}{6 \pi \eta a}\approx \frac{0.2}{a}\,\,\,\,\, $ & $\displaystyle \approx 20\, \frac{\mu \mathrm{m}^2}{\mathrm{s}} $ &  $\displaystyle \approx 0.2\, \frac{\mu \mathrm{m}^2}{\mathrm{s}}$ \\ \noalign{\medskip} \hline \hline

Hydrodynamic numbers &  $a=10 n \mathrm{m}$ & $a=1 \mu \mathrm{m}$ \\  
\noalign{\medskip}
\hline \noalign{\medskip}

Pe $\displaystyle = \frac{v_S a}{D_{\mathrm{col}}} \approx 5 v_S \, a^2$ &
$\displaystyle\approx 0.005$ & $\displaystyle\approx 50$ \\
\noalign{\medskip}

Re $\displaystyle= \frac{v_S a}{\nu} \approx 10^{-6} v_S a$ & 
$\displaystyle\approx 10^{-7}$ &  $\displaystyle\approx 10^{-5}$ \\
\noalign{\medskip}

Kn $\displaystyle =\frac{\lambda_\mathrm{free}}{a} \approx \frac{0.3}{a} \times 10^{-3}$ &
 $\displaystyle \approx 0.03$ & $\displaystyle
\approx 0.0003$ \\ \noalign{\medskip}

Ma $\displaystyle =\frac{v_S}{c_s} \approx 6.76 \times 10^{-10} v_S $ 
& $\approx 6.8 \times 10^{-10} $ 
& $\approx 6.8 \times 10^{-10} $  \\ \noalign{\bigskip}

\hline \hline

Time-scales &  $a=10 n \mathrm{m}$ & $a= 1 \mu \mathrm{m}$ \\ \noalign{\medskip} \hline \noalign{\medskip}

$\displaystyle\tau_D = \frac{a^2}{D_{\mathrm{col}}} \approx 5 a^3 \, $&
$\displaystyle \approx 5 \times 10^{-6} \mathrm{s} $ & $\displaystyle \approx 5 \,\,
 \mathrm{s}$ \\ \noalign{\medskip}

$\displaystyle t_S = \frac{a}{v_S} = \frac{\tau_\nu}{Re} =
\frac{\tau_{D}}{Pe} $& $\displaystyle = 0.001\,\, \mathrm{s}$ &
$\displaystyle = 0.1 \,\,\mathrm{s}$ \\ \noalign{\medskip}

$\displaystyle\tau_\nu = \frac{a^2}{\nu} \approx 10^{-6} a^2 \, $&
$\displaystyle \approx 10^{-10} \mathrm{s}$ & $\displaystyle
\approx 10^{-6} \mathrm{s}$ \\\noalign{\medskip}

$\displaystyle\tau_B = \frac{M}{\xi_S} = \frac{2}{9} \tau_\nu \, $&
$\displaystyle \approx 2.2 \times 10^{-11} \mathrm{s}$ &
$\displaystyle  \approx 2.2 \times 10^{-7} \mathrm{s}$
\\\noalign{\medskip}

$\displaystyle t_{cs} = \frac{a}{c_s} \approx 6.7 \times 10^{-10} a \, $&
$\displaystyle \approx 6.7 \times 10^{-12} \mathrm{s} $ &
$\displaystyle  \approx 6.7 \times 10^{-10} \mathrm{s} $
\\\noalign{\medskip}

\end{tabular}
\end{ruledtabular}
\end{center}
\end{table}

\begin{table}
\begin{center}
\caption{Physical parameters, dimensionless hydrodynamic numbers and
time-scales for a stick boundary colloid in an SRD fluid.
In the first column, we take the same small $\lambda$ limit as in Table~\ref{tableSRD}  to highlight the dominant scaling.
In the other columns more accurate values for $\nu/\nu_0$ are used. The hydrodynamic numbers and times were estimated
with a length-scale $a\approx\sigma_{cf}\approx \sigma_{cc}/2$. 
 The colloids are neutrally buoyant,
$\rho_c = \rho_f = \gamma m_f /a_0^3$, and their velocity is
$v_S=0.05a_0/t_0$, i.e.\ Ma=$0.039$. Units  as in 
Table~\protect\ref{tableSRD}.
\label{tablecolloidSRD}}
\begin{ruledtabular}
\begin{tabular}{lll}
Physical parameters &  $\sigma_{cf}=2 \, a_0$ & $\sigma_{cf}= 10 \, a_0$ \\  \noalign{\medskip}\hline \noalign{\medskip}

$\displaystyle M  \approx \frac{4}{3} \pi  \gamma \sigma_{cf}^3  \frac{m_f}{a_0^3} \,  $& $\displaystyle
 \approx  168 m_f $  & $\displaystyle  2.09 \times 10^4 m_f$  \\\noalign{\medskip}

$\displaystyle \frac{\xi_E}{\xi_S}  \approx 8.6 \,\, \sigma_{cf} \lambda \,  $& $\displaystyle
 \approx  1.8  $  & $\displaystyle  \approx 9.0 $  \\\noalign{\medskip}

$\displaystyle \xi_S = 6 \pi \eta a \approx \frac{\gamma \sigma_{cf}}{\lambda}  $&
$\displaystyle \approx  96 \,\, \frac{m_f}{t_0} $ &
$\displaystyle \approx 478 \frac{m_f}{t_0} \,\,  $
\\\noalign{\medskip}

$\displaystyle D_{\mathrm{col}} \approx  \frac{k_B T}{6 \pi \eta a}\approx \frac{\lambda}{\gamma \sigma_{cf}} \,\,\,\,\, $ & $\displaystyle \approx 0.0165 \frac{a_0^2}{t_0} \,\,\,\,\,\,$ &  $\displaystyle \approx 0.00236\, \frac{a_0^2}{t_0}$ \\ \noalign{\medskip} \hline\hline

Hydrodynamic numbers &  $\sigma_{cf}=2 \, a_0$ & $\sigma_{cf}= 10 \, a_0 $ \\ \noalign{\medskip} \hline \noalign{\medskip}

Pe $\displaystyle = \frac{v_S a}{D_{\mathrm{col}}} \approx 1.29 \mbox{Ma}\frac{\gamma  \sigma_{cf}^2}{ \lambda}$ &
$\displaystyle\approx 6.2 $ & $\displaystyle\approx  212 $ \\
\noalign{\medskip}

Re $\displaystyle= \frac{v_S a}{\nu} \approx  23\,\, \mbox{Ma} \, \lambda \sigma_{cf} $ & 
$\displaystyle\approx 0.2 $ &  $\displaystyle \approx 1$ \\
\noalign{\medskip}

Kn $\displaystyle =\frac{\lambda_\mathrm{free}}{a} \approx \frac{\lambda}{\sigma_{cf}}$ &
 $\displaystyle \approx 0.05$ & $\displaystyle
\approx 0.01$ \\ \noalign{\medskip}

Ma $\displaystyle =\frac{v_S}{c_s}  $ & $\approx 0.039$
& $\approx 0.039$ \\ \noalign{\medskip}

\hline\hline
Time-scales &  $\sigma_{cf} = 2 \, a_0$ & $\sigma_{cf} = 10 \, a_0$ \\ \noalign{\medskip} \hline \noalign{\medskip}

$\displaystyle\tau_D = \frac{a^2}{D_{\mathrm{col}}} \approx  \frac{\gamma \sigma_{cf}^3}{\lambda}  $&
$\displaystyle \approx 242 t_0$ & $\displaystyle \approx 4.2
\times 10^{4} \, t_0$ \\ \noalign{\medskip}

$\displaystyle t_S = \frac{a}{v_s} = \frac{\tau_\nu}{Re} =
\frac{\tau_{D}}{Pe}\approx \frac{\sigma_{cf}}{1.29 \mbox{Ma}} $& $\displaystyle \approx 31 t_0$ &
$\displaystyle  \approx 155 t_0$ \\ \noalign{\medskip}

$\displaystyle\tau_\nu = \frac{a^2}{\nu} \approx 18 \, \lambda \sigma_{cf}^2  \, $&
$\displaystyle \approx  8.0 t_0 $ & $\displaystyle
\approx   200 t_0$ \\\noalign{\medskip}

$\displaystyle\tau_B = \frac{M}{\xi_S} \approx \frac{2}{9} \tau_v  $&
$\displaystyle  \approx 1.75 t_0 $ &
$\displaystyle  \approx  44 t_0$ \\\noalign{\medskip}

$\displaystyle\tau_E = \frac{M}{\xi_E} \approx 0.5 \sigma_{cf}  $&
$\displaystyle  \approx 0.98 t_0 $ &
$\displaystyle  \approx  4.9 t_0$
\\\noalign{\medskip}

$\displaystyle t_{cs} = \frac{\sigma_{cf}}{c_s} \approx 0.78 \sigma_{cf}   $&
$\displaystyle  \approx 1.55 t_0 $ &
$\displaystyle  \approx  7.8  t_0$
\\\noalign{\medskip}

\end{tabular}
\end{ruledtabular}
\end{center}

\end{table}

\FloatBarrier


\begin{thebibliography}{99}
\bibitem{BH} Note that one could also define the volume fraction
using an effective hard-sphere radius derived from the
Barker-Henderson or other related criteria~\protect\cite{Hans86}.
\end{thebibliography}
\end{document}